\documentclass[reprint,aps,graphicx,pre,showpacs]{revtex4-1}
\usepackage{amsmath,amssymb,amsfonts}     

\usepackage{graphicx,bm}
  \usepackage{paralist}
  \usepackage{epstopdf}
  \usepackage{graphics} 
 \usepackage[colorlinks=true]{hyperref}
 \hypersetup{urlcolor=blue, citecolor=red}
 \usepackage[latin1]{inputenc}
\usepackage[caption=false]{subfig}

\newcommand{\dis}{\displaystyle}
 \usepackage{tikz-cd}

\newcommand{\calP}{{\mathcal P}}

\newcommand{\calF}{{\mathcal F}}

\newcommand{\calM}{{\mathcal M}}
\newcommand{\calC}{{\mathcal C}}

\newcommand{\R}{{\mathbb R}}
\renewcommand{\L}{{\mathbb L}}
\newcommand{\X}{\mathbf{X}}
\renewcommand{\P}{\mathbb{P}}
\newcommand{\PP}{\widetilde{P}}
\newcommand{\Q}{\widetilde{Q}}
\newcommand{\x}{\mathbf{x}}
\newcommand{\y}{\mathbf{y}}
\newcommand{\z}{\mathbf{z}}
\newcommand{\e}{{\mathrm e}}
\newcommand{\E}{{\mathbb E}}

\newcommand{\n}{\mathbf n}
\newcommand{\calT}{{\mathcal T}}
\renewcommand{\P}{\mathbb P}
\newcommand{\p}{\widetilde{p}}

\newcommand{\q}{\widetilde{q}}

\begin{document}

 \title{Renewal equations for single-particle diffusion through a semipermeable interface}

\author{Paul C. Bressloff$^{1}$}
\address{Department of Mathematics, University of Utah 155 South 1400 East, Salt Lake City, UT 84112}
%%%% Subject entries to be placed here %%%%

%%%% Keyword entries to be placed here %%%%

%%%% Insert corresponding author and its email address}
\date{\today}

\begin{abstract} 
Diffusion through semipermeable interfaces has a wide range of applications, ranging from molecular transport through biological membranes to reverse osmosis for water purification using artificial membranes. At the single-particle level, one-dimensional diffusion through a barrier with constant permeability $\kappa_0$ can be modeled in terms of so-called snapping out Brownian motion (BM). The latter sews together successive rounds of partially reflecting BMs that are restricted to either the left or right of the barrier.  Each round is killed (absorbed) at the barrier when its Brownian local time exceeds an exponential random variable parameterized by $\kappa_0$. A new round is then immediately started in either direction with equal probability. It has recently been shown that the probability density for snapping out BM satisfies a renewal equation that relates the full density to the probability densities of partially reflected BM on either side of the barrier. Moreover, generalized versions of the renewal equation can be constructed that incorporate non-Markovian, encounter-based models of absorption. In this paper we extend the renewal theory of snapping out BM to single-particle diffusion in bounded domains and higher spatial dimensions. In each case we show how the solution of the renewal equation satisfies the classical diffusion equation with a permeable boundary condition at the interface. That is, the probability flux across the interface is continuous and proportional to the difference in densities on either side of the interface. We also consider an example of an asymmetric interface in which the directional switching after each absorption event is biased. Finally, we show how to incorporate an encounter-based model of absorption for single-particle diffusion through a spherically symmetric interface. We find that, even when the same non-Markovian model of absorption applies on either side of the interface, the resulting permeability is an asymmetric time-dependent function with memory. Moreover, the permeability functions tend to be heavy-tailed.

\end{abstract}
%%%%%%%%%%%%%%%%%%%%%%%%%%%

\maketitle

\section{Introduction}

\begin{figure}[t!]
  \centering
  \includegraphics[width=6cm]{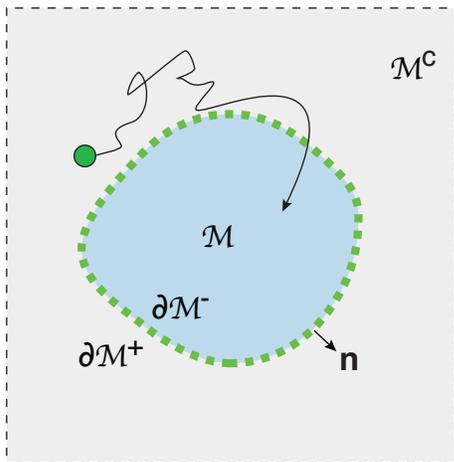}
  \caption{Diffusion through a closed semipermeable membrane in $\R^d$.}
  \label{fig1}
\end{figure}

A classical problem in the theory of diffusion is transport through a semipermeable interface. For example, suppose that $\calM$ denotes a closed bounded domain $\calM\subset \R^d$ with a smooth concave boundary $\partial \calM$ separating the two open domains $\calM$ and its complement $ \calM^c$, see Fig. \ref{fig1}.  The boundary acts as a semipermeable interface with $\partial \calM^+ $ ($\partial \calM^-$) denoting the side approached from outside (inside) $\calM$, see Fig. \ref{fig1}. Let $u(\x,t)$ be the concentration of particles at $\x$ at time $t$. Then $u(\x,t)$ is the weak solution of the diffusion equation with a permeable or leather boundary condition on $\partial \calM$
\begin{subequations}
\label{dclass}
\begin{align}
\frac{\partial u(\x,t)}{\partial t}&=D\nabla^2u(\x,t) ,\ \x \in\calM\cup \calM^c,\\
J(\y^{\pm},t)&=\kappa_0[u(\y^-,t)-u(\y^+,t)],\quad \y^{\pm} \in \partial \calM^{\pm},
\end{align}
\end{subequations}
where $J(\x,t)=-D\nabla u(\x,t) \cdot \n$ is the particle flux, $\n$ is the unit normal directed out of $\calM$, $D$ is the diffusivity and $\kappa_0$ is the (constant) permeability. Eqs. (\ref{dclass}) are a special case of the well-known Kedem-Katchalsky (KK) equations \cite{Kedem58,Kedem62,Kargol96}, which also allow for discontinuities in the diffusivity and chemical potential across the interface. The macroscopic KK equations can be derived by considering a thin membrane and using statistical thermodynamics. More simply, Eqs. (\ref{dclass}) arise from treating the interface as a thin layer of slow diffusion $D=O(h)$, where $h$ is the width of the layer, and taking the limit $h\rightarrow 0$ \cite{Aho16}.
Although the KK equations were originally developed within the context of the transport of non-electrolytes through biological membranes, they are now used to describe all types of membranes, both biological and artificial. (See the recent collection of articles in Ref. \cite{Nik21}.) One application of artificial membranes is reverse osmosis for water purification and for extracting energy from variations in salinity \cite{Li10,Rubinstein21}.

The macroscopic theory of diffusion through semipermeable membranes has motivated a number of stochastic models at the single-particle level. One approach is to consider random walks on lattices in which semipermeable barriers are represented by local defects \cite{Powles92,Kenkre08,Novikov11,Kay22}. 
An alternative approach is to use stochastic differential equations (SDEs). These generate sample paths of a Brownian particle that are distributed according to a probability density satisfying a corresponding FP equation. However, incorporating the microscopic analog of the permeable boundary condition (\ref{dclass}b) is non-trivial. If $\partial \calM$ were a totally reflecting (Neumann) or partially reflecting (Robin) boundary, then Brownian motion (BM) confined to $\calM$ would need to be supplemented by an additional impulsive force each time the particle contacted the boundary (prior to possible absorption). Mathematically speaking, this can be implemented by introducing a Brownian functional known as the boundary local time \cite{Ito65,Freidlin85,Papanicolaou90,Milshtein95,Borodin96,Grebenkov06}. The latter determines the amount of time that a Brownian particle spends in the neighborhood of points on the boundary. A rigorous probabilistic formulation of one-dimensional BM in the presence of a semipermeable barrier is much more recent. It is based on so-called snapping out BM, which was first introduced by Lejay \cite{Lejay16}, see also Refs. \cite{Aho16,Lejay18,Brobowski21}. Snapping out BM sews together successive rounds of partially reflecting BM that are restricted to either $x<0$ or $x>0$ with a semipermeable barrier at $x=0$. Suppose that the particle starts in the domain $x>0$. It realizes positively reflected BM until its local time exceeds an exponential random variable with parameter $\kappa_0$. It then immediately resumes either negatively or positively reflected BM with equal probability, and so on. Snapping out BM is itself a generalization of so-called skew BM \cite{Ito63}, which has a wide range of applications, particularly in mathematical finance \cite{Lejay06,Decamps06,App11,Gairat17}. (Note that  SDEs in the form of underdamped Langevin equations have been used to develop efficient computational schemes for finding solutions to the FP equation in the presence of one or more semipermeable interfaces \cite{Farago18,Farago20}. This is distinct from snapping out BM, which is an exact single-particle realization of diffusion through an interface in the overdamped limit.)

We recently reformulated snapping out BM in terms of a renewal equation that related the full probability density to the probability densities of the partially reflected BMs on either side of the barrier \cite{Bressloff22p}.  (The original analysis of Lejay \cite{Lejay16} used the theory of semigroups and resolvent operators to derive a corresponding backward equation.)  The renewal equation was solved using Laplace transforms and Green's function methods, resulting in an explicit expression for the probability density of snapping out BM. We then used the renewal approach to develop a more general probabilistic model of one-dimensional single-particle diffusion through a semi-permeable barrier. This included modifications of the diffusion process away from the barrier such as stochastic resetting \cite{Evans20}, and encounter-based models of membrane absorption \cite{Grebenkov20,Grebenkov22,Bressloff22,Bressloff22a} that kill each round of partially reflected BM. In the latter case, the corresponding boundary condition at the interface involved a time-dependent permeability with memory.  

In this paper we extend the renewal theory of snapping out BM to single-particle diffusion in bounded domains and higher spatial dimensions. We  first consider the example of a bounded interval partitioned by a semipermeable membrane, and with a reflecting  boundary at each end. We then turn to a higher-dimensional version of snapping out BM which corresponds to the configuration shown in Fig. \ref{fig1}. In both cases we show how the solution of the renewal equation satisfies a FP equation of the form (\ref{dclass}). Establishing such an equivalence is non-trivial, since one needs to take into account modifications in the partially reflecting boundary conditions when the particle starts exactly on the boundary. (This is related to the notion of the so-called inverse local time \cite{Ito65}). Although one could proceed by directly solving the corresponding FP equation (\ref{dclass}), the renewal approach has at least two potential advantages. First, since snapping out BM generates sample paths of single-particle diffusion through semipermeable interfaces, it can be used to develop numerical schemes for generating solutions to the corresponding FP, see also \cite{Farago18,Farago20}. Second, the renewal equation provides a framework for developing more general probabilistic models along the lines considered in \cite{Bressloff22p}.

The structure of the paper is as follows. In section II we construct the renewal equations for snapping out BM in an interval with reflecting external boundaries and a semipermeable barrier within the interior. We show that the probability density satisfies the FP equation with a permeable boundary condition at the barrier. We then extend the analysis to the case of an asymmetric interface in which the directional switching after each absorption event is biased. We also consider a first passage time (FPT) problem for an asymmetric barrier and a right-hand boundary that is totally absorbing. We show that the mean FPT (MFPT) is independent of the permeability $\kappa_0$ if the particle starts to the right of the barrier, but there is a jump in the MFPT and its first derivative with respect to the initial position as the latter crosses the barrier. In section III we consider the renewal equation for a closed semipermeable membrane in $\R^d$, and show that the probability density satisfies an FP equation of the form (\ref{dclass}). We then explicitly solve the renewal equation for a spherically symmetric interface. Finally, in Section IV we incorporate an encounter-based model of absorption into the spherically symmetric example. In particular, we show that non-Markovian models of absorption generate an asymmetric time-dependent permeability distribution that tends to be heavy-tailed. The emergence of time-dependent kernels through non-Markovian absorption is analogous to a recent study of thin membrane boundary conditions based on a random walk model with nonexponential waiting times within the membrane \cite{Kos21}.

\vfill

  \setcounter{equation}{0}
\section{Snapping out BM in an interval}

\begin{figure}[h!]
  \centering
  \includegraphics[width=8cm]{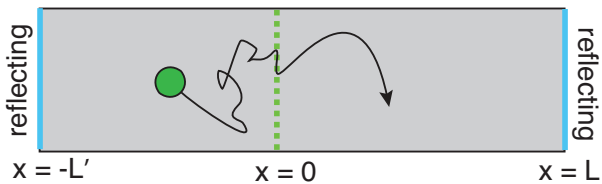}
  \caption{Brownian motion in the interval $[-L',L]$ with a semipermeable membrane at $x=0$ and reflecting boundary conditions at $x=-L',L$.}
  \label{fig2}
\end{figure}

Consider a Brownian particle diffusing in the interval $[-L',L]$ with a semipermeable barrier at $x=0$ and reflecting boundaries at the ends $x=-L',L$, see Fig. \ref{fig2}.
Introduce the disjoint sets $[0^+,L]$ and $[-L',0^-]$ with $0^{\pm}$ denoting the position of the barrier when approaching from either the left-hand or right-hand sides.
Let $\rho(x,t|x_0)$ denote the probability density of the particle position under the initial condition $X_0=x_0 \neq 0^{\pm}$ and set 
\begin{equation}
\rho(x,t)=\int_{-L'}^{L}\rho(x,t|x_0)g(x_0)dx_0
\end{equation} 
for any continuous function $g$ such that $\int_{-L'}^{L}g(x_0)dx_0=1$. The classical way to determine $\rho(x,t)$ would be to solve the corresponding FP equation
\begin{subequations}
\label{interval}
\begin{align}
&\frac{\partial \rho}{\partial t}=D\frac{\partial^2\rho}{\partial x^2}, \quad x\in (-L',0^-)\cup (0^+,L),\\
&\left . D\frac{\partial \rho(x,t)}{\partial x}\right |_{x=0^{\pm}}=\kappa_0 [\rho(0^+,t)-\rho(0^-,t)],\\
&\left . D\frac{\partial \rho(x,t)}{\partial x}\right |_{x=-L',L}=0.
\end{align}
\end{subequations}
In this section we follow a different approach by constructing a renewal equation that relates $\rho(x,t)$ to the probability densities of partially reflected BM in the two intervals $[-L',0]$ and $[0,L]$, respectively. This generalizes the construction presented in Ref. \cite{Bressloff22p} for snapping out BM in $\R$. Using a combination of Green's function methods and Laplace transforms, we establish that the solution of the renewal equation satisfies Eqs. (\ref{interval}). 
Hence, analytically solving the FP equation reduces to the problem of calculating the Green's functions for partially reflected BM in an interval. 

\subsection{Green's function for partially reflected BM}

Consider BM in the interval $[0,L]$ with $x=0$ partially reflecting and $x=L$ totally reflecting. Let $X_t\in [0,L]$ denote the position of the Brownian particle at time $t$ and introduce the Brownian local time
\begin{equation}
\label{loc}
\ell_t=\lim_{\epsilon\rightarrow 0} \frac{D}{\epsilon} \int_0^tH(\epsilon-X_{\tau})d\tau,
\end{equation}
where $H$ is the Heaviside function. Note that $\ell_t$, which has units of length due to the additional factor of $D$, determines the amount of time that the Brownian particle spends in the neighborhood of $x=0$ over the interval $[0,t]$. It can be shown that $\ell_t$ exists and is a nondecreasing, continuous function of $t$ \cite{Ito65}. The partially reflecting boundary condition at $x=0$ can be implemented by introducing the stopping time
\begin{equation}
\calT=\inf\{t>0: \ell_t>\widehat{\ell}\},\quad \P[\widehat{\ell}>\ell]\equiv \Psi(\ell)=\e^{-\kappa_0\ell/D}.
\end{equation}
That is the stochastic process is killed when the local time exceeds a random exponentially distributed threshold. The probability density for particle position prior to absorption at $x=0$ \cite{Ito65,Freidlin85,Papanicolaou90,Milshtein95,Borodin96,Grebenkov06},
\begin{equation}
p(x,t|x_0)dx=\P[x\leq X_t<x+dx, t<\calT|X_0=x_0],
\end{equation}
satisfies the FP equation with a Robin boundary condition at $x=0$:
  \begin{subequations}
\label{Robin2}
\begin{align}
\frac{\partial p(x,t|x_0)}{\partial t}&=D\frac{\partial^2 p(x,t|x_0)}{\partial x^2}, \quad 0<x<L,\\
D\partial_xp(0,t|x_0)&=\kappa_0 p(0,t|x_0),\ -D\partial_xp(L,t|x_0)=0,
\end{align}
\end{subequations}
and $p(x,0|x_0)=\delta(x-x_0)$. It is convenient to Laplace transform with respect to $t$, which gives 
  \begin{subequations}
\label{RobinLT}
\begin{align}
&D\frac{\partial^2\widetilde{p}(x,s|x_0)}{\partial x^2}-s \widetilde{p}(x,ts|x_0)=-\delta(x-x_0),\\
&D\partial_x\widetilde{p}(0,s|x_0)=\kappa_0 \widetilde{p}(0,s|x_0),\\ &-D\partial_x\p(L,s|x_0)=0,
\end{align}
\end{subequations}
with $0<x,x_0<L$.
We can identify $\widetilde{p}(x,s|x_0)$ as a Green's function of the modified Helmholtz equation on $[0,L]$. The general solution for $0<x<x_0$, after imposing the Robin boundary condition at $x=0$, is proportional to the density
\begin{align}
\widetilde{p}_<(x,s)&= \frac{1}{2}\left [\e^{\sqrt{s/D}x}+\frac{\sqrt{sD}-\kappa_0}{\sqrt{sD}+\kappa_0}\e^{-\sqrt{s/D}x}\right ]   \\
&=\frac{\sqrt{sD}\cosh(\sqrt{s/D}x)+\kappa_0\sinh(\sqrt{s/D}x)}{\sqrt{sD}+\kappa_0}.\nonumber
\end{align}
Similarly, the solution for $x_0<x<L$, which satisfies the reflecting boundary condition at $x=L$, is of the form\begin{equation}
\widetilde{p}_>(x,s)= \cosh(\sqrt{s/D}(L-x)).
\end{equation}
Imposing continuity of $\p(x,s|x_0)$ across $x_0$ and matching the discontinuity in the first derivative yields the solution
 \begin{align}
 \widetilde{p}(x,s|x_0)= \left \{ \begin{array}{cc} A \widetilde{p}_<(x,s)\widetilde{p}_>(x_0,s), & 0\leq x\leq x_0\\ & \\
 A\widetilde{p}_>(x,s)\widetilde{p}_<(x_0,s), & x_0\leq x\leq L\end{array}
 \right .
 \label{solVN}
 \end{align}
 with
 \begin{equation}
  A=A(\kappa_0,s)\equiv \frac{(\sqrt{sD}+\kappa_0)/\sqrt{sD}} {\sqrt{sD}\sinh(\sqrt{sD}L)+\kappa_0\cosh(\sqrt{s/D}L)}.
  \end{equation}
 In particular, note that
 \begin{equation}
 \widetilde{p}(x,s|0)=\frac{\cosh(\sqrt{s/D}(L-x))}{\sqrt{sD}\sinh(\sqrt{s/D}L)+\kappa_0\cosh(\sqrt{s/D}L)},
 \end{equation}
and
 \begin{equation}
 \label{kol}
 D\partial_x\widetilde{p}(0,s|0)=\kappa_0\widetilde{p}(0,s|0)-1.
 \end{equation}
 The modification of the Robin boundary condition when the particle starts at the barrier plays a significant tole in establishing the equivalence of snapping out BM.

Note that the boundary condition (\ref{kol}) when $x_0=0$ is a modified version of the Robin boundary condition when $x_0>0$. 
Moreover, the Green's function $\widetilde{p}(x,s|0)$ can be related to the so-called inverse local time \cite{Ito65}. The latter is defined according to
\begin{equation}
\E[\e^{-s \calT}|X_0=x_0]=\int_0^{\infty} f(x_0,t) \e^{-s\calT} dt,
\end{equation}
where $f(x_0,t)$ is the FPT density. In terms of the survival probability
\begin{equation}
Q(x_0,t)=\int_0^{\infty}p(x,t|x_0)dx,
\end{equation}
we have
\begin{align}
f(x_0,t)&=-\frac{dQ(x_0,t)}{dt}=-\int_0^{\infty}\frac{\partial p(x,t|x_0)}{\partial t}dx\nonumber \\
&=-D\int_0^{\infty}\frac{\partial^2 p(x,t|x_0)}{\partial x^2}dx= \left .D\frac{\partial p(x,t|x_0)}{\partial x}\right |_{x=0}\nonumber \\
&= \kappa_0 p(0,t|x_0).
\end{align}
Hence,
\begin{equation}
\E[\e^{-s\calT}|X_0=x_0]=\kappa_0 \widetilde{p}(0,s|x_0)=\kappa_0\widetilde{p}(x_0,s|0).
\end{equation}
We have used the well-known symmetry property of the Green's function for a self-adjoint operator.

\subsection{Renewal equation for snapping out BM}

We construct snapping out BM in $[-L',L]$ as follows \cite{Lejay16}. Without loss of generality, assume that the particle starts at $X_0=x_0\geq 0$. It realizes positively reflected BM until its local time $\ell_t$ at $x=0^+$ is greater than an independent exponential random variable $\widehat{\ell}$ of parameter $\kappa_0$. Let $\calT_0$ denote the corresponding stopping time. The process immediately restarts as a new reflected BM with probability 1/2 in either $[0^+,L]$ or $[-L',0^-]$ and a reset local time. Again the reflected BM is stopped when the reset local time exceeds a new exponential random variable etc. 
Let $p(x,t|x_0)$ and $q(x,t|x_0)$ denote the probability densities of partially reflected BM in the intervals $[0^+,L]$ and $[0^+,L']$, respectively, and set
\begin{align}
p(x,t)&=\int_0^Lp(x,t|x_0)g(x_0)dx_0,\ x\in [0^+,L]\\
q(x,t)&= \int_{-L'}^0q(-x,t|x_0)g(x_0)dx_0,\ x\in [-L',0^-].
\end{align}
In particular, the Laplace transform $\q(x,s|x_0)$ is given by Eq. (\ref{solVN}) under the mapping $L\rightarrow L'$. Since snapping out BM satisfies the strong Markov property \cite{note1}, as previously shown by Lejay \cite{Lejay16}, there exists a last renewal equation analogous to the one introduced in Ref. \cite{Bressloff22p}:
\begin{subequations}
\label{renewal}
 \begin{align}
  & \rho(x,t)=p(x,t)\\
&\qquad +\frac{\kappa_0}{2} \int_0^t p(x,\tau|0)[\rho(0^+,t-\tau )+\rho(0^-,t-\tau )]d\tau \nonumber
    \end{align}
    for $x\in [0^+,L]$ and
     \begin{align}
& \rho(x,t)=q(x,t)\\
&\qquad +\frac{\kappa_0}{2} \int_0^t q(-x,\tau|0)[\rho(0^+,t-\tau )+\rho(0^-,t-\tau )]d\tau \nonumber
    \end{align}
    \end{subequations}
    for $x\in [-L',0^-]$. The first term on the right-hand side of Eq. (\ref{renewal}a) represents all sample trajectories that have never been absorbed by the barrier at $x=0^{+}$ up to time $t$. The integral in Eq. (\ref{renewal}a) sums over all trajectories that were last absorbed (stopped) at time $t-\tau$ in either the positively or negatively reflected BM state and then switched with probability 1/2 to the positive side in order to reach $x$ at time $t$. Since the particle is not absorbed over the interval $(t-\tau,t]$, the probability of reaching $x$ is $p(x,\tau|0)$. The terms in Eq. (\ref{renewal}b) have the corresponding interpretations in $[-L',0^-]$. Finally, the probability that the last stopping event occurred in the interval $(t-\tau,t-\tau+d\tau)$ irrespective of previous events is $\kappa_0 d\tau$. 
    
Clearly $\rho(x,t)$ satisfies the diffusion equation in the bulk, so we will focus on the boundary conditions at the semipermeable barrier. It is convenient to Laplace transform the renewal Eqs. (\ref{renewal}) with respect to time $t$ by setting $\widetilde{\rho}(x,s) =\int_0^{\infty}\e^{-st}\rho(x,t)dt$ etc. This gives
 \begin{subequations}
 \label{renewal2}
 \begin{align}
 \widetilde{\rho}(x,s) &= \p(x,s)+ \frac{\kappa_0}{2} \p(x,s|0)\Sigma_{\rho}(s),\, x\in [0^+,L], \\
  \widetilde{\rho}(x,s) &= \q(x,s)+ \frac{\kappa_0}{2} \q(-x,s|0)\Sigma_{\rho}(s),\,x\in [-L',0^-]
 \end{align}
 \end{subequations}
 where
 \begin{equation}
 \Sigma_{\rho}(s)=\widetilde{\rho}(0^+,s )+\widetilde{\rho}(0^-,s ) .
 \end{equation}
 Setting $x=0+$ and $x=0^-$ in Eqs. (\ref{renewal2}a,b), respectively, summing the results and rearranging shows that 
  \begin{align}
  \label{Lam0}
\Sigma_{\rho}(s) = \frac{ \Sigma_p(s)}{1- \kappa_0[\p(0,s|0)+\q(0,s|0)]/2} ,
 \end{align}
 where $\Sigma_p(s)=\p(0^+,s)+\q(0^-,s)$.

  Next, differentiating Eqs. (\ref{renewal2}a,b) with respect to $x$ and setting $x=0^{\pm}$ gives
  \begin{subequations}
  \label{pho}
  \begin{align}
\partial_x \widetilde{\rho}(0^+,s)&=\partial_x\p(0^+,s)+\frac{\kappa_0}{2}\partial_x\p(0,s|0)\Sigma_{\rho}(s),\\
\partial_x \widetilde{\rho}(0^-,s)&=\partial_x\q(0^-,s)-\frac{\kappa_0}{2}
 \partial_x\q(0,s|0)\Sigma_{\rho}(s) .
  \end{align}
  \end{subequations}
Imposing the Robin boundary condition (\ref{RobinLT}b) implies that
\[D\partial_x\p(0^+,s)=\kappa_0 \p(0^+,s),\quad D\partial_x\q(0^-,s)=-\kappa_0 \q(0^-,s).\]
 On the other hand, Eq. (\ref{kol}) yields
 \begin{align*}
 D\partial_x\p(0,s|0)&=\kappa_0\p(0,s|0)-1,\\  D\partial_x\q(0,s|0)&=\kappa_0\q(0,s|0)-1.
 \end{align*}
 Substituting into Eqs. (\ref{pho}a,b), we have
   \begin{subequations}
  \label{pho2}
  \begin{align}
D\partial_x \widetilde{\rho}(0^+,s)&=\kappa_0\p(0^+,s)+\frac{\kappa_0}{2}[\kappa_0\p(0,s|0)-1]\Sigma_{\rho}(s) ,\\
D\partial_x \widetilde{\rho}(0^-,s)&=-\kappa_0\q(0^-,s)-\frac{\kappa_0}{2}
 [\kappa_0\q(0,s|0)-1]\Sigma_{\rho}(s) .
  \end{align}
  \end{subequations}
Subtracting Eqs. (\ref{pho2}a,b) and using Eq. (\ref{Lam0}) implies that
   \begin{align}
&D[\partial_x \widetilde{\rho}(0^+,s)-\partial_x \widetilde{\rho}(0^-,s)] =\kappa_0\Sigma_p(s) \nonumber\\
&+ \kappa_0 \{\kappa_0[\p(0,s|0)+\q(0,s|0)]/2-1\}\Sigma_{\rho}(s)=0.
 \label{dev1}
  \end{align}
Similarly, adding equations (\ref{pho2}a,b),
  \begin{align}
  2 D\partial_x \widetilde{\rho}(0^{\pm},s)&=\kappa_0[\p(0^+,s)-\q(0^-,s)]\nonumber \\
  &\quad +\frac{\kappa_0^2}{2}[\p(0,s|0)-\q(0,s|0)]\Sigma_p(s) \nonumber \\
&=\kappa_0[\widetilde{\rho}(0^+,s)-\widetilde{\rho}(0^-,s)].
\label{dev2}
\end{align}
 
 Eqs. (\ref{dev1}) and (\ref{dev2}) establish that the density $\widetilde{\rho}(x,s)$ satisfies the Laplace transform of the semipermeable membrane BVP (\ref{interval}) under the initial condition $\rho(x,0)=g(x)$ and $\kappa_0\rightarrow \kappa_0/2$. Hence, the snapping out BM $X_t$ on ${\mathbb G}$ is the single-particle realization of the stochastic process whose probability density evolves according to the diffusion equation with a semipermeable membrane at $x=0$. In the symmetric case $L'=L$ with $g(x_0)$ an even function of $x_0$, we find that $\widetilde{\rho}(x,s)$ is an even function of $x$ so that the flux through the membrane is zero. In other words, it effectively acts as a totally reflecting barrier even though $\kappa_0>0$. It can also be checked that the solution of Eq. (\ref{renewal2}) reduces to
\begin{equation}
\widetilde{\rho}(x,s)=\frac{1}{4\sqrt{sD}}\left (\e^{-\sqrt{s/D}|x-x_0|}+\e^{-\sqrt{s/D}(x+x_0)}\right )
\end{equation} 
for $x>0$. Finally, note that we recover the results of Ref. \cite{Bressloff22p} in the limits $L,L'\rightarrow \infty$.

\subsection{Snapping out BM with imperfect contacts}
A classical generalization of the permeable boundary condition (\ref{interval}b) is to include a directional asymmetry in the permeability, which can be interpreted as a step discontinuity in a chemical potential \cite{Kedem58,Kedem62,Kargol96,Farago20}:
\begin{equation}
-D \partial_x u(0^+,t)=-D  \partial_x u(0^-,t) = \kappa_0[u(x^+,t)-\sigma u(x^-,t)]
\end{equation}
for $0\leq \sigma \leq 1$. This tends to enhance the concentration to the left of the barrier. (If $\sigma >1$ then we would have a barrier with permeability $\kappa_0\sigma$ and bias $1/\sigma$ to the right. Here we show how to incorporate the directional asymmetry into snapping out BM. The basic idea is to consider a bias in the switching between the positive and negative directions of reflected BM following each round of killing. More specifically, consider the transitions
\begin{equation}
\label{switch}
0^{\pm}\overset{\alpha\kappa_0}\rightarrow 0^{+},\quad 0^{\pm}\overset{\beta\kappa_0}\rightarrow 0^{-},\quad \alpha+\beta=1.
\end{equation}
The renewal equation (\ref{renewal2}) becomes
\begin{subequations}
  \label{bias2}
\begin{align}
 \widetilde{\rho}(x,s) &= \p(x,s)+\alpha \kappa_0 \p(x,s|0)\Sigma_{\rho}(s)
 \end{align}
 for $x\in [0^+,L]$ and
 \begin{align}
 \widetilde{\rho}(x,s) &= \q(x,s)+\beta \kappa_0 \q(-x,s|0)\Sigma_{\rho}(s) \end{align}
 \end{subequations}
 for $x\in [-L',0^-]$
 
 Setting $x=0^{\pm}$ in equations (\ref{bias2}), summing the results and rearranging yields the explicit solution
\begin{equation}
\label{Lam1}
\Sigma_{\rho}(s)=
 \frac{ \p(0^+,s)+\q(0^-,s)}{1-\kappa_0[\alpha \p(0,s|0)+\beta\q(0,s|0)]} .
 \end{equation}
Using a similar argument to the unbiased case, we obtain the pair of equations
\begin{subequations}
  \label{phoal}
  \begin{align}
\partial_x \widetilde{\rho}(0^+,s)&=\kappa_0\p(0^+,s)+\alpha \kappa_0[\kappa_0\p(0,s|0)-1]\Sigma_{\rho}(s),\\
\partial_x \widetilde{\rho}(0^-,s)&=-\kappa_0\q(0^-,s)-\beta \kappa_0
 [\kappa_0\q(0,s|0)-1]\Sigma_{\rho}(s).
  \end{align}
  \end{subequations}
Subtracting this pair of equations and using (\ref{Lam1}) with $\alpha+\beta=1$ establishes that the flux is continuous across the membrane.
 On the other hand, multiplying Eq. (\ref{phoal}a) by $\beta$ and Eq. (\ref{phoal}b) by $\alpha$, and adding the results yields
 \begin{align}
2D \partial_x\widetilde{\rho}(0^{\pm},s)    &= \kappa_0[\beta \p(0^+,s)-\alpha \p(0^-,s)]\nonumber \\
&\quad +\alpha \beta \kappa_0^2[\p(0,s|0)-\q(0,s|0)]\Sigma_{\rho}(s)\nonumber \\
&= \kappa_0 \left [ \beta\widetilde{\rho}(0^+,s)- \alpha \widetilde{\rho}(0^-,s) \right ].
  \end{align}

 Hence, snapping out BM with the switching scheme (\ref{switch}) and $\alpha < \beta$ is equivalent to single-particle diffusion through a directed semipermeable barrier with an effective permeability $\kappa_0\beta/2$ and bias $\sigma =\alpha/\beta$ on the left-hand side. Similarly, when $\alpha >\beta$, we have a directed semipermeable barrier with an effective permeability $\kappa_0\alpha/2$ and bias $\sigma =\beta/\alpha$ on the right-hand side.
 
 \subsection{First-passage time problem}

As a simple application of the renewal Eq. (\ref{phoal}), consider the FPT problem obtained by replacing the reflecting boundary at $x=L$ in Fig. \ref{fig2} by a totally absorbing boundary. The only modification to our previous analysis is that the Laplace transformed probability density in the domain $[0,L]$ is now given by Eq. (\ref{solVN}) with
\begin{subequations}
\begin{align}
&\widetilde{p}_<(x,s)=\frac{\sqrt{sD}\cosh(\sqrt{s/D}x)+\kappa_0\sinh(\sqrt{s/D}x)}{\sqrt{sD}+\kappa_0},\\
&\widetilde{p}_>(x,s)= \sinh(\sqrt{s/D}(L-x)),\\
 A&=\frac{(\sqrt{sD}+\kappa_0)/\sqrt{sD}} {\sqrt{sD}\cosh(\sqrt{sD}L)+\kappa_0\sinh(\sqrt{s/D}L)}.
\end{align}
\end{subequations}
Let $\calT_L$ denote the FPT to be absorbed at $x=L$,
\begin{equation}
\calT_L=\inf\{t>0, X_t=L\},
\end{equation}
Take $f(x_0,t)$ to be the FPT density when $X_0=x_0$. We identify $f(x_0,t)$ with the flux through $x=L$, $f(x_0,t)=-D\partial_x\rho(L,t|x_0)$. It follows that the MFPT is
\begin{align}
\E[\calT_L]\equiv \int_0^{\infty} tf(t)dt&=-\left . \frac{d\widetilde{J}(x_0,s)}{ds}\right |_{s=0}\nonumber \\
&=\left .D\frac{d}{ds}\partial_x\widetilde{\rho}(L,s|x_0)\right|_{s=0}.
\end{align}

 \begin{figure}[b!]
  \centering
  \includegraphics[width=7cm]{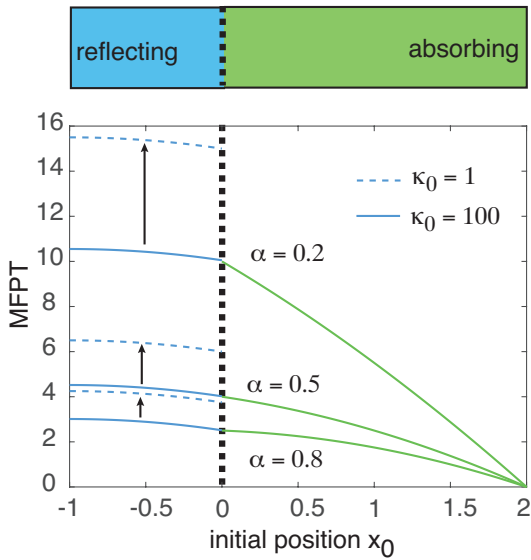}
  \caption{Asymmetric semipermeable barrier at $x=0$ with a reflecting boundary at $x=L'=-1$ and an absorbing boundary at $x=L=2$. Plots of MFPT $\E[\calT_L]$ as function of the initial position $x_0$ for various $\alpha$ and $\kappa_0$. We also set $D=1$.}
  \label{fig3}
\end{figure}

First, suppose that $x_0 <0$. From Eqs. (\ref{phoal}a) and (\ref{Lam1}), we have
\begin{align}
 \partial_x\widetilde{\rho}(L,s|x_0) &= \frac{  \alpha\kappa_0\partial_x\p(L,s|0)\q(0,s |-x_0)}{1- \kappa_0[\alpha\p(0,s|0)+(1-\alpha)\q(0,s|0)]},\nonumber
   \end{align}
  where
 \begin{equation}
 \widetilde{p}(x,s|0)=\frac{\sinh(\sqrt{s/D}(L-x))}{\sqrt{sD}\cosh(\sqrt{s/D}L)+\kappa_0\sinh(\sqrt{s/D}L)},
 \end{equation}
and
 \begin{equation}
 \widetilde{q}(x,s|0)=\frac{\cosh(\sqrt{s/D}(L'-x))}{\sqrt{sD}\sinh(\sqrt{s/D}L')+\kappa_0\cosh(\sqrt{s/D}L')}.
 \end{equation}
We find that for $\kappa_0>0$
\begin{equation}
\label{MFPTm}
\E[\calT_L]=\frac{(L+L')^2}{2D}-\frac{(L'+x_0)^2}{2D}+\frac{LL'(1-2\alpha)}{\alpha D}+\frac{L'}{\alpha \kappa_0}.
\end{equation}
Now suppose that $x_0>0$. In this case we have 
\begin{align}
 \partial_x\widetilde{\rho}(L,s|x_0) &=\partial_x \p(L,s |x_0)  \\
 &\quad + \frac{\alpha  \kappa_0 \partial_x\p(L,s|0)\p(0,s |x_0)}{1- \kappa_0[\alpha\p(0,s|0)+(1-\alpha)\q(0,s|0)]}.\nonumber
   \end{align}
and after some algebra we find that for $\kappa_0>0$,
\begin{align}
 \label{MFPTp}
 \E[\calT_L]&=\frac{(L+L')^2}{2D}-\frac{(L'+x_0)^2}{2D}\nonumber \\
 &\quad + \frac{L'(L-x_0)}{D}  \frac{1-2\alpha}{\alpha}.
 \end{align}
  Eqs. (\ref{MFPTm}) and (\ref{MFPTp}) generalize the recent result for the symmetric case $\alpha=1/2$, which was obtained by solving a backward equation for the MFPT:
 \begin{equation}
 \E[\calT_L]=\frac{(L+L')^2}{2D}-\frac{(L'+x_0)^2}{2D}+\frac{2L'}{  \kappa_0}H(-x_0)
 \end{equation}
 for $x_0\in [-L',L]$.

A number of observations can be made. First, if the particle starts to the right of the barrier, then the MFPT is independent of the permeability $\kappa_0$ for all $\alpha \in [0,1]$. As $\kappa_0$ increases, there is a higher probability of crossing the barrier to the left-hand side, but it is also easier for the particle to cross back to the right-hand side; these effects cancel out. As highlighted in Ref. \cite{Kay22}, this is a consequence of the fact that diffusion is unbiased. Second, the MFPT is a continuous function of $x_0$ across the barrier in the limit $\kappa_0 \rightarrow \infty$, whereas its first derivative is discontinuous (unless $\alpha=1/2$). Third, there is an additional contribution to the MFPT for $x_0<0$ given by $L'/(\alpha \kappa_0)$, which 
represents the mean time to cross the barrier for the first time. Fourth, the MFPT is a decreasing function of $\alpha$ for all $x_0$. Example plots of the MFPT as a function of $x_0$ is illustrated in Fig. \ref{fig3} for various values of $\alpha$ and $\kappa_0$. Finally, note that the limit $\kappa_0\rightarrow 0$ is singular since $\E[\calT_L]$ does not exist for $x_0<0$ and $\E[\calT_L]=L^2/2D-x_0^2/2D$ for $x_0>0$. 

\begin{figure*}[t!]
\includegraphics[width=14cm]{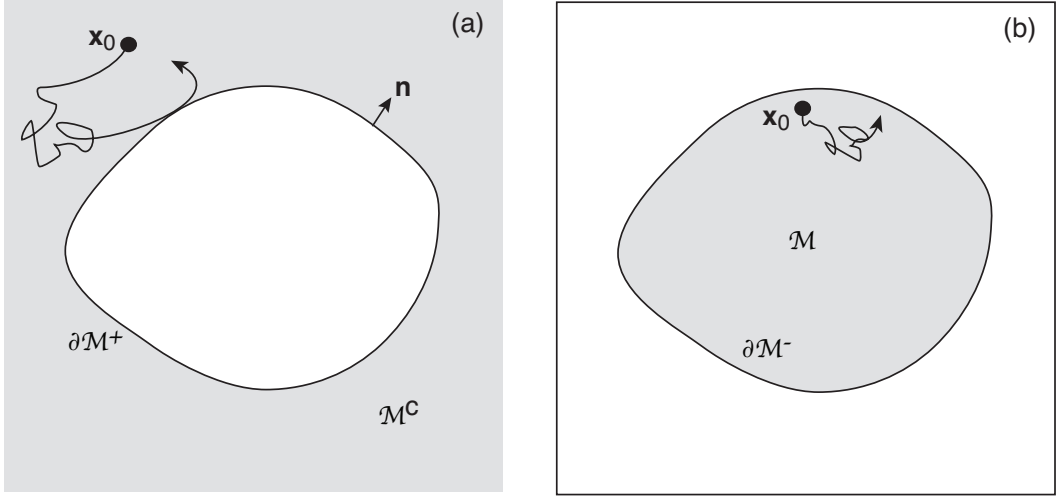}
\caption{Decomposition of a higher-dimensional snapping out BM into two partially reflected BMs corresponding to (a) $\X_t\in \calM^c$ and (b) $\X_t\in \calM$ , respectively.}
\label{fig4}
\end{figure*}

\setcounter{equation}{0}   
\section{Snapping out BM in $\R^d$}

Let us return to the setup of Fig. \ref{fig1}. Single-particle diffusion now takes place on the space ${\mathbb G}=\overline{\calM}\cup \overline{\calM^c}$. Here $\overline{\calM}=\calM\cup \partial \calM^-$ and $\overline{\calM^c}=\calM^c\cup \partial \calM^+$ are disjoint sets so that $\y \in \partial \calM$ corresponds to either $\y^+\in \partial \calM^+$ or $\y^- \in \partial \calM^-$ treated as distinct points. Let $\rho(\x,t|\x_0)$, $\x,\x_0\in {\mathbb G}$, denote the probability density of the particle with the initial condition $\X_0=\x_0\in  {\calM}\cup  {\calM^c}$ and set 
\begin{equation}
\rho(\x,t)=\int_{\mathbb G} \rho(\x,t|\x_0)g(\x_0)d\x_0
\end{equation} 
for any continuous function $g$ on ${\mathbb G}$ with $\int_{\mathbb G}g(\x_0)d\x_0=1$. The density $\rho$ satisfies the FP equation 
\begin{subequations}
\label{Rd}
\begin{align}
\frac{\partial \rho(\x,t)}{\partial t}&=D\nabla^2\rho(\x,t) ,\ \x \in\calM\cup \calM^c,\\
J(\y^{\pm},t)&=\kappa_0[\rho(\y^-,t)-\rho(\y^+,t)],\quad \y^{\pm} \in \partial \calM^{\pm},
\end{align}
\end{subequations}
together with the initial condition $\rho(\x,0) =g(\x)$. We wish to derive the higher-dimensional version of the renewal equations (\ref{renewal}) by sewing together partially reflected BMs in the domains $\calM$ and $\calM^c$, see Fig. \ref{fig4}.

\subsection{Partially reflected BMs in $\calM$ and $\calM^c$}

Consider a Brownian particle diffusing in the bounded domain $ \calM$, see Fig. \ref{fig4}(a) with $\partial \calM^-$ totally reflecting. Let $\X_t$ denote the position of the particle at time $t$. In order to write down a stochastic differential equation (SDE) for $\X_t$, we introduce the boundary local time
\begin{equation}
\label{loc}
\ell_t^-=\lim_{\epsilon\rightarrow 0} \frac{D}{\epsilon} \int_0^tH(\epsilon-\mbox{dist}(\X_{\tau},\partial \calM^-))d\tau,
\end{equation}
such that the corresponding SDE takes the form 
\begin{equation}
d\X_t =\sqrt{2D}d{\bf W}_t-\n(\X_t)  d\ell_t^-,
\end{equation}
where ${\bf W}_t$ is a $d$-dimensional Brownian motion and $\n(\X_t)$ is the outward unit normal at the point $\X_t \in \partial \calM$. The differential $d\ell^-_t$ can be expressed in terms of a Dirac delta function:
\begin{equation}
d\ell^-_t=Ddt\left (\int_{\partial \calM^-}\delta(\X_t-\y)d\y\right ) .
\end{equation}
Partially reflected BM iin $\calM$ is then obtained by stopping the stochastic process $\X_t$ when the local time $\ell_t^-$ exceeds a random exponentially distributed threshold $\widehat{\ell}$ \cite{Grebenkov06}. That is, the particle is absorbed somewhere on $\partial \calM^-$ at the stopping time
 \begin{equation}
\label{exp}
{\mathcal T}^-=\inf\{t>0:\ \ell_t^- >\widehat{\ell}\},\quad \P[\widehat{\ell}>\ell]  =\e^{-\kappa_0 \ell/D}.
\end{equation}
The marginal density for particle position (prior to absorption), 
\[q(\x,t|\x_0)dx=\P[\x\leq \X_t < \x+d\x ,\ t < {\mathcal T}^-|\X_0=\x_0],  \]
satisfies the diffusion equation with a Robin boundary condition on $\partial \calM^-$:
\begin{subequations} 
\label{qdiffloc}
\begin{align}
	&\frac{\partial q(\x,t|\x_0)}{\partial t} = D\nabla^2 q(\x,t|\x_0) \mbox{ for } \x,\x_0 \in  {\calM},\\
&D\nabla q(\x,t|\x_0) \cdot \n=-\kappa_0q(\x,t|\x_0) \mbox{ for } \x\in \partial \calM^-,
\label{qdiffloc2}
	\end{align}
	\end{subequations}
and $q(\x,0|\x_0)=\delta(\x-\x_0)$.	
	
	An analogous construction holds for partially reflected BM in $ \calM^c$, see Fig. \ref{fig4}(b). Given the local time
\begin{equation}
\label{locp}
\ell_t^+=\lim_{\epsilon\rightarrow 0} \frac{D}{\epsilon} \int_0^tH(\epsilon-\mbox{dist}(\X_{\tau},\partial \calM^+))d\tau,
\end{equation}
and stopping time
\begin{equation}
\label{expp}
{\mathcal T}^+=\inf\{t>0:\ \ell_t^+ >\widehat{\ell}\},\quad \P[\widehat{\ell}>\ell] =\e^{-\kappa_0 \ell/D}.
\end{equation}
one finds that the marginal density
\[p(\x,t|\x_0)dx=\P[\x\leq \X_t < \x+d\x ,\ t < {\mathcal T}^+|\X_0=\x_0]\]
satisfies the Robin  boundary value problem (BVP)
\begin{subequations} 
\label{pdiffloc}
\begin{align}
	&\frac{\partial p(\x,t|\x_0)}{\partial t} = D\nabla^2 p(\x,t|\x_0) \mbox{ for } \x,\x_0 \in  {\calM^c},\\
&D\nabla p(\x,t|\x_0) \cdot \n=\kappa_0p(\x,t|\x_0) \mbox{ for } \x\in \partial \calM^+,
	\end{align}
	\end{subequations} 
and $p(\x,0|\x_0)=\delta(\x-\x_0)$.

\subsection{Modified boundary condition for $\x_0\in \partial \calM$}

\begin{figure}[t!]
  \centering
  \includegraphics[width=6cm]{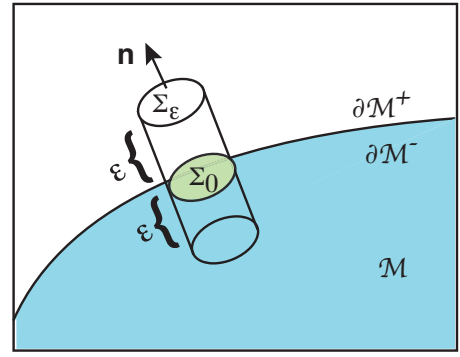}
  \caption{Cylinder construction across the semipermeable membrane. See text for details.}
  \label{fig5}
\end{figure}

As in the 1D case, the boundary condition for partially reflected BM in $\calM^c$ is modified when the particle actually starts on the boundary. In order to show this,  we first Laplace transform Eqs. (\ref{pdiffloc}) with respect to time $t$:
\begin{subequations} 
\label{pdifflocLT}
\begin{align}
	& D\nabla^2 \p(\x,s|\x_0) -s\p(\x,s|\x_0)=-\delta(\x-\x_0),\, \x,\x_0 \in  {\calM^c},\\
&D\nabla \p(\x,s|\x_0) \cdot \n=\kappa_0\p(\x,s|\x_0) \mbox{ for } \x\in \partial \calM^+.
	\end{align}
	\end{subequations} 
Consider a small cylinder ${\mathcal C}(\epsilon,\sigma)$ of uniform cross-section $\sigma$ and length $2\epsilon$ with a point $\y \in \partial \calM$ at its center of mass, see Fig. \ref{fig5}. Let ${\mathcal C}^+(\epsilon,\sigma)={\mathcal C}(\epsilon,\sigma)\cap\overline{ \calM^c}$ For sufficiently small $\sigma$, we can treat $\Sigma_{0}\equiv {\mathcal C}^+(\epsilon,\sigma)\cap \partial \calM^+$ as a planar interface with outward normal $\n(\y)$ such that the axis of ${\mathcal C}^+(\epsilon,\sigma)$ is aligned along $\n(\y)$. Given the above construction, we integrate Eq. (\ref{pdifflocLT}a) with respect to all $\x\in  {\mathcal C}^+(\epsilon,\sigma)$ and use the divergence theorem:
 \begin{align}
& \int_{\Sigma_{\epsilon}} \nabla \p(\y',s|\x_0)\cdot \n(\y') d\y' -\int_{\Sigma_{0}} \nabla \p(\y',s|\x_0) \cdot \n(\y') d\y' \nonumber \\
&\quad \sim \frac{1}{D}\int_{\calC^+} [s\p(\x,s|\x_0)-\delta(\x-\x_0)] d\x,
\label{cyl}
 \end{align}
 where $\Sigma_{\epsilon}$ denotes the flat end of the cylinder within $\calM^c$. If $\x_0$ is in the bulk domain $\calM^c$, then taking the limits $\epsilon,\sigma \rightarrow 0$ shows that the flux is continuous as it approaches the boundary, since the right-hand side of Eq. (\ref{cyl}) vanishes. On the other hand, if $\x_0=\z \in\partial \calM^+$ then taking the limits $\epsilon,\sigma \rightarrow 0$ gives
 \begin{align}
 &\lim_{\epsilon \rightarrow 0^+} D \nabla \p(\y+\epsilon \n(\y),s|\z)\cdot \n(y) \nonumber \\
 &\quad  -D  \nabla \p(\y,s|\z) \cdot \n(\y)=-\overline{\delta}(\y-\z),
 \end{align}
 where $\overline{\delta}$ is the Dirac delta function for points on $\partial \calM$ such that for any continuous function $f: \calM\rightarrow \R$ we have
 $\int_{\partial \calM} f(\y)\overline{\delta}(\y-\z)d\y = f(\z)$. Finally, noting that the first flux term on the left-hand side satisfies the boundary condition (\ref{pdifflocLT}b), we deduce that
 \begin{equation}
 \label{crucial}
 D  \nabla \p(\y,s|\z) \cdot \n(\y)=\kappa_0 \p(\y,s|\z)-\overline{\delta}(\y-\z).
 \end{equation}
 Applying a similar argument to partially reflected BM in $\calM$ we find that
  \begin{equation}
 \label{crucial2}
 D  \nabla \q(\y,s|\z) \cdot \n(\y)=-\kappa_0 \q(\y,s|\z)+\overline{\delta}(\y-\z).
 \end{equation}
 The extra terms on the right-hand side of Eqs. (\ref{crucial}) and (\ref{crucial2}) play a crucial role in the subsequent analysis. They will also be confirmed by directly differentiating example explicit solutions.
 
 The Green's function $\p(\x_0,s|\z)$ with $\z \in \partial \calM$ and $\x_0 \in  {\calM^c}$ can be related to the corresponding inverse local time \cite{Ito65}
\begin{equation}
\label{survd}
\E[\e^{-s \calT^+}|\X_0=\x_0]=\int_0^{\infty} f(\x_0,t) \e^{-st} dt,
\end{equation}
where $f(\x_0,t)$ is the FPT density for being absorbed on $\partial \calM$. In terms of the survival probability
\begin{equation}
Q(\x_0,t)=\int_{\calM^c}p(\x,t|\x_0)d\x,
\end{equation}
we have
\begin{align}
&f(\x_0,t)=-\frac{dQ(\x_0,t)}{dt}=-\int_{\calM^c}\frac{\partial p(\x,t|\x_0)}{\partial t}d\x\nonumber \\
&=-D\int_{\calM^c} \nabla^2 p(\x,t|\x_0)d\x \\
&=  D\int_{\partial \calM} \nabla p(\z,t|\x_0)\cdot \n d\z= \kappa_0 \int_{\partial \calM} p(\z,t|\x_0) d\z.\nonumber
\end{align}
Hence,
\begin{align}
\E[\e^{-s\calT}|\X_0=\x_0]&=\kappa_0\int_{\partial \calM} \p(\z,s|\x_0) d\z\nonumber\\
&=\kappa_0\int_{\partial \calM} \p(\x_0,s|\z) d\z
\end{align}
by the standard symmetry property of Green's functions.

\subsection{Renewal equation}

We define the multidimensional version of snapping out BM as follows. 
Without loss of generality, suppose that the particle starts in the domain $ {\calM^c}$. It realizes reflected BM in ${\calM^c}$ until it is killed when its local time $\ell_t^+$, see Eq. (\ref{locp}), is greater than an independent exponential random variable $\widehat{\ell}$. Let $\y^+\in \partial \calM^+$ denote the point on the boundary where killing occurs. The stochastic process immediately restarts as a new round of partially reflected BM, either from $\y^+$ into $\calM^c$ or from $\y^-$ into $\calM$. These two possibilities occur with equal probability. Subsequent rounds of partially reflected BM are generated in the same way. We thus have a stochastic process on the set ${\mathbb G}$. As in the one-dimensional case \cite{Lejay16}, it can be proven that snapping out BM is a strong Markov process. This means that we can consider a multi-dimensional version of the renewal equation introduced in \cite{Bressloff22}. First, let 
\begin{subequations}
 \begin{align}
 p(\x,t)&=\int_{\overline{\calM^c}}p(\x,t|\x_0)g(\x_0)d\x_0,\\ q(\x,t)&=\int_{\overline{\calM}}q(\x,t|\x_0)g(\x_0)d\x_0,
 \end{align}
 \end{subequations}
 where $p(\x,t|\x_0)$ and $q(\x,t|\x_0) $ are the solutions of the Robin BVPs (\ref{pdiffloc}) and (\ref{qdiffloc}), respectively. By construction, the probability density $\rho(\x,t)$ satisfies the last renewal equations
 \begin{widetext}
  \begin{subequations}
  \label{drenewal}
\begin{align}
  \rho(\x,t)&=p(\x,t)+\frac{\kappa_0}{2}\int_0^t \left \{ \int_{\partial \calM}p(\x,\tau|\z)[\rho(\z^+,t-\tau ) +\rho(\z^-,t-\tau )]d\z\right \}d\tau ,\quad  \x\in \overline{\calM^c} ,\\
  \rho(\x,t)&=q(\x,t)+\frac{\kappa_0}{2}\int_0^t \left \{ \int_{\partial \calM}q(\x,\tau|\z)[\rho(\z^+,t-\tau ) +\rho(\z^-,t-\tau )]d\z\right \}d\tau , \quad  \x\in  \overline{\calM }.
   \end{align}
 \end{subequations} 
 \end{widetext}
  The first term on the right-hand side of Eqs. (\ref{drenewal}a) and (\ref{drenewal}b) represents all sample trajectories that have never been absorbed by the boundary $\partial \calM^+$ and $\partial \calM^-$, respectively. The corresponding integral term in equation (\ref{drenewal}a) represents all trajectories that were last absorbed (stopped) somewhere on $\partial \calM^{\pm}$ at time $t-\tau$ and then switched to the domain $\overline{\calM^c}$ with probability 1/2 in order to reach $\x\in\overline{\calM^c}$ at time $t$. Since the particle is not absorbed over the interval $(t-\tau,t]$, the probability of reaching $\x \in \overline{\calM^c}$ starting at a point $\z\in \partial \calM^+$ is $p(\x,\tau|\z)$. We then have to integrate with respect to all starting positions  $\z$ at time $t-\tau$. An analogous interpretation holds for the integral term on the right-hand side of Eq. (\ref{drenewal}b), with $p\rightarrow q$ and $\partial \calM^+\rightarrow \partial \calM^-$. Finally, the probability that the last stopping event occurred in the interval $(t-\tau,t-\tau+d\tau)$ irrespective of previous events is $\kappa_0 d\tau$. 

We wish to establish that $\rho(\x,t)$ is a (weak) solution of the FP Eq. (\ref{Rd}) under the initial condition $\rho(\x,0)=g(\x)$. It is clear that $\rho(\x,t)$ satisfies the diffusion equation in the bulk so, as in the 1D example, we focus on the boundary conditions. Laplace transforming the renewal Eqs. (\ref{drenewal}a,b) with respect to time $t$ gives
\begin{subequations}
 \label{mrenewal}
 \begin{align}
  \widetilde{\rho}(\x,s) = \p(\x,s) +\frac{\kappa_0}{2} \int_{\partial \calM}\p(\x,s|\z)\Sigma_{\rho}(\z,s)d\z 
  \end{align}
  for $\x\in \overline{\calM^c}$ and
  \begin{align}
  \widetilde{\rho}(\x,s) = \q(\x,s)  +\frac{\kappa_0}{2} \int_{\partial \calM}\q(\x,s|\z)\Sigma_{\rho}(\z,s)d\z 
 \end{align} 
 \end{subequations}
for $  \x\in \overline{\calM}$. We have set
\begin{equation}
\Sigma_{\rho}(\z,s)=\widetilde{\rho}(\z^+,s )+\widetilde{\rho}(\z^-,s ) .
\end{equation}
Taking the normal derivative of Eqs. (\ref{mrenewal}a,b) with $\partial_{\n}\equiv \n \cdot \nabla$ in the limit $\x\rightarrow \y \in \partial \calM$ gives the pair of equations
 \begin{subequations}
 \label{mrenewal2}
 \begin{align}
 \partial_{\n} \widetilde{\rho}(\y^+,s) &= \partial_{\n} \p(\y^+,s)  +\frac{\kappa_0}{2} \int_{\partial \calM}\partial_{\n}\p(\y,s|\z)\Sigma_{\rho}(\z,s)d\z, \\
 \partial_{\n} \widetilde{\rho}(\y^-,s) &= \partial_{\n}\q(\y^-,s) +\frac{\kappa_0}{2} \int_{\partial \calM}\partial_{\n}\q(\y,s|\z)\Sigma_{\rho}(\z,s)d\z .
 \end{align} 
 \end{subequations}
 
 Next, imposing the boundary conditions (\ref{pdifflocLT}b) and (\ref{pdifflocLT}b) for partially reflected BM and the modified boundary conditions (\ref{crucial}) and (\ref{crucial2}) yields
 \begin{subequations}
 \label{mrenewal3}
 \begin{align}
 D\partial_{\n} \widetilde{\rho}(\y^+,s) &=  \kappa_0  \p(\y^+,s)\\
  &  +\frac{\kappa_0}{2} \int_{\partial \calM}[\kappa_0\p(\y,s|\z)-\overline{\delta}(\y -\z)]\Sigma_{\rho}(\z,s)d\z,\nonumber \\
 D\partial_{\n} \widetilde{\rho}(\y^-,s) &=-\kappa_0\q(\y^-,s)\\
  &-\frac{\kappa_0}{2} \int_{\partial \calM}[\kappa_0\q(\y,s|\z)-\overline{\delta}(\y-\z)]\Sigma_{\rho}(\z,s)d\z. \nonumber
 \end{align} 
 \end{subequations}
 Subtracting this pair of equations, we find that
  \begin{align}
 &D\partial_{\n} \widetilde{\rho}(\y^+,s) -D\partial_{\n} \widetilde{\rho}(\y^-,s) \nonumber\\
 &= \kappa_0 [\p(\y^+,s)+\q(\y^-,s)]-\kappa_0\Sigma_{\rho}(\y,s)\nonumber  \\
  &  \quad +\frac{\kappa_0^2}{2} \int_{\partial \calM}[ \p(\y,s|\z)+\q(\y,s|\z)]\Sigma_{\rho}(\z,s)d\z  =0.
 \end{align} 
 The last line follows from setting $\x=\y^+$ and $\x=\y^-$ in Eqs. (\ref{mrenewal}a) and (\ref{mrenewal}b), respectively, and adding the results.
 Finally adding Eqs. (\ref{mrenewal3}a,b) gives
\begin{align}
 2D\partial_{\n} \widetilde{\rho}(\y^{\pm},s)&= \kappa_0 [\p(\y^+,s)-\q(\y^-,s)]\nonumber  \\
  &  +\frac{\kappa_0^2}{2} \int_{\partial \calM}[ \p(\y,s|\z)-\q(\y,s|\z)]\Sigma_{\rho}(\z,s)d\z\nonumber \\
& =\kappa_0[\widetilde{\rho}(\y^+,s )-\widetilde{\rho}(\y^-,s ) ].
 \end{align} 
Hence, we have established the equivalence of multidimensional snapping out BM with single-particle diffusion through a smooth semipermeable membrane of the form shown in Fig. \ref{fig1}.

 \subsection{Spectral decomposition}
 
 Eqs. (\ref{mrenewal2}) are Fredholm integral equations of the second kind for which $\rho$ is an implicit solution. One way to formally solve these equations is to use spectral theory. Setting $\x=\y^{\pm}$ and adding the resulting equations gives
\begin{align}
\Sigma_{\rho}(\y,s)&=\Sigma_p(\y,s)+\frac{\kappa_0}{2} \int_{\partial \calM}\Sigma_p(\y,s|\z) \Sigma_{\rho}(\z,s)d\z,\nonumber \\
&\quad \mbox{for } \y \in \partial \calM,
\label{sig1}
\end{align}
with
 \begin{subequations}
\begin{align}
\Sigma_p(\y,s|\z)&=\p(\y,s|\z )+\q(\y,s|\z ) ,\\ \Sigma_p(\y,s)&=\int_{\mathbb G} \Sigma_p(\y,s|\x_0)g(\x_0)d\x_0.
\end{align}
 \end{subequations}
Introduce the linear operator ${\mathbb L}:\partial \calM \rightarrow \partial \calM$,
\begin{equation}
 {\mathbb L}[f](\y,s)=\int_{\partial \calM} \Sigma_p(\y,s|\z)f(\z)d\z,
 \end{equation}
 for any $L^2$ function $f$ on $\partial \calM$
 and rewrite Eq. (\ref{sig1}) as
 \begin{equation}
 \label{sig2}
 \Sigma_{\rho}(\y,s)-\frac{\kappa_0}{2}{\mathbb L}[\Sigma_{\rho}](\y,s)=\Sigma_p(\y,s).
 \end{equation}
Since $\partial \calM$ is bounded and ${\mathbb L}$ is self-adjoint with respect to the $\L^2$ norm, it follows that ${\mathbb L}$ has a complete orthonormal set of eigenfunctions $\{\phi_n(\y,s), n\geq 0\}$ and a corresponding set of real nonzero eigenvalues $\lambda_n(s)$ such that
 \begin{equation}
 {\mathbb L}\phi_n(\y,s)=\lambda_n(s)\phi_n(\y,s),\ \y \in \partial \calM.
 \end{equation}
 
 Introducing the eigenfunction expansions
  \begin{subequations}
 \begin{align}
 \Sigma_{\rho}(\y,s)&=\sum_{n\geq 0}\Sigma_{\rho,n}(s)\phi_n(\y,s),\\
   \Sigma_{p}(\y,s)&=\sum_{n\geq 0}\Sigma_{p,n}(s)\phi_n(\y,s).
 \end{align}
 \end{subequations}
and substituting into Eq. (\ref{sig2}) yields
 \begin{equation}
 \Sigma_{\rho,n}(s) =\frac{ \Sigma_{p,n}(s) }{1-\kappa_0\lambda_n(s)/2}.
 \end{equation}
We have used the fact that the eigenfunctions are orthonormal. Finally, substituting for $\Sigma_{\rho}(\y,s)$, $\y \in \partial \calM$, in Eqs. (\ref{mrenewal2}) gives 
 \begin{subequations}
 \begin{align}
 \label{pLTrenewala}
  \widetilde{\rho}(\x,s) &= \p(\x,s) \\
  &\quad +\frac{\kappa_0}{2}\sum_{n\geq 0} \frac{ \Sigma_{p,n}(s) }{1-\kappa_0\lambda_n(s)/2} \int_{\partial \calM}\p(\x,s|\z)\phi_n(\z,s)d\z \nonumber  \end{align}
  for $\x\in \overline{\calM^c}
$, and
  \begin{align}
  \label{pLTrenewalb}
  \widetilde{\rho}(\x,s) &= \q(\x,s) \\
  & +\frac{\kappa_0}{2}\sum_{n\geq 0} \frac{ \Sigma_{p,n}(s) }{1-\kappa_0\lambda_n(s)/2}\int_{\partial \calM}\q(\x,s|\z)\phi_n(\z,s)d\z \nonumber
 \end{align} 
 for $ x\in \overline{\calM}$.
 \end{subequations}

 \subsection{Spherically symmetric semipermeable interface}

In special cases, it is possible to solve the renewal Eqs. (\ref{mrenewal2}) without recourse to spectral theory by exploiting an underlying symmetry. For example, suppose that $\calM =\{\x\in \R^d\,|\, 0\leq  |\x| <R\}$ and thus $\partial \calM =\{\x\in \R^d\,|\,  |\x| =R\}$, where $R$ is the radius of the sphere.
Following \cite{Redner01}, we assume that the initial distribution of the particle is spherically symmetric, that is, $g=g(|\x_0|)$. This allows us to exploit spherical symmetry by setting 
\begin{align}
\rho=\rho(r,t)= {\Omega_d} \int_0^{\infty} \rho(r,t|r_0)g(r_0)r_0^{d - 1} dr_0,
\end{align}
where $r= |{\bf x}|$, $r_0=|\x_0|$ and $\Omega_d$ is the surface area of a unit sphere in $\mathbb{R}^d$. The renewal Eqs. (\ref{mrenewal}) reduce to the simpler form
\begin{subequations}
 \label{sprenewal}
 \begin{align}
  \widetilde{\rho}(r,s) &= \p(r,s) +\frac{\kappa_0}{2}\Omega_d R^{d-1} \p(r,s|R) \Sigma_{\rho}(R,s) 
  \end{align}
  for $r\geq R^+$ and
  \begin{align}
  \widetilde{\rho}(r,s) &= \q(r,s) 
   +\frac{\kappa_0}{2}\Omega_d R^{d-1}\q(r,s|R) \Sigma_{\rho}(R,s) 
 \end{align} 
 \end{subequations}
for $0\leq r\leq R^-$. We have also set
\begin{equation}
\Sigma_{\rho}(R,s)=\widetilde{\rho}(R^+,s )+\widetilde{\rho}(R^-,s ).
\end{equation}
 Eqs. (\ref{sprenewal}) are identical in structure to Eqs. (\ref{renewal2}). This means that we can immediately write down the solution for $\Sigma_{\rho}(R,s)$:  \begin{align}
  \label{spLam0}
\Sigma_{\rho}(R,s) = \frac{ \Sigma_p(R,s)}{1- ({\kappa(R)}/{2})[\p(R,s|R)+\q(R,s|R)]} ,
 \end{align}
 with $\kappa(R)=\kappa_0\Omega_d R^{d-1}$.
 Hence, obtaining an explicit solution for $ \widetilde{\rho}(r,s)$ reduces to the problem of solving the FP equation for $\p(r,s|r_0)$, $r,r_0\geq R$, and the corresponding FP equation for $\q(r,s|r_0)$, $0\leq r,r_0\leq R$.
 
The Laplace transformed density $\p(r,s|r_0)$ satisfies the FP equation
\begin{subequations}
\label{sph}
\begin{align}
 &D\frac{\partial^2\p(r,s|r_0)}{\partial r^2} + D\frac{d - 1}{r}\frac{\partial \p(r,s|r_0)}{\partial r}-s\p(r,s|r_0)\nonumber \\
 &\hspace{3cm} =-\Gamma_d\delta(r-r_0) ,\quad R <r,\\
  &D\frac{\partial \p(r,s|r_0)}{\partial r}=\kappa_0 \p(r,s|r_0) ,\quad r=R,
\end{align}
\end{subequations}
with $\Gamma_d=1/(\Omega_dr_0^{d-1})$.
Equations of the form (\ref{sph}) can be solved in terms of modified Bessel functions \cite{Redner01}. The general solution is
  \begin{eqnarray}
    \p(r, s|r_0) = A(s)F(\eta r)+ G(r, s|r_0), \, \eta=\sqrt{\frac{s}{D}}  
    \label{qir}
\end{eqnarray}
for $R\leq r$, where $F(x)=x^{\nu}K_{\nu}(x)$, $\nu = 1 - d/2$, and $K_{\nu}$ is a modified Bessel function of the second kind.
The first term on the right-hand side of Eq. (\ref{qir}) is the solution to the homogeneous version of Eq. (\ref{sph}) and $G$ is the modified Helmholtz Green's function in the case of a totally absorbing surface $\partial \calM$:
  \begin{subequations}
\begin{align}
\label{Ga}
&D\frac{\partial^2G}{\partial r^2} + D\frac{d - 1}{r}\frac{\partial G}{\partial r} -sG  = -\Gamma_d\delta(r - r_0), \, R<r,\\ 
 & G(R,s|r_0)=0.
 \label{Gb}
\end{align}
\end{subequations}
The latter is given by \cite{Redner01}
\begin{align}
\label{GGs}
 &  G(r, s| r_0) = \frac{ (rr_0)^\nu }{D\Omega_d}\\
   &\qquad \times \frac{[I_{\nu}(\eta r_<)K_{\nu}(\eta R)-I_{\nu}(\eta R)K_{\nu}(\eta r_<)]K_{\nu}(\eta r_>)}{K_{\nu}(\eta R)},\nonumber
\end{align}
where $r_< = \min{(r, r_0)}$, $r_> = \max{(r, r_0)}$, and $I_{\nu}$ is a modified Bessel function of the first kind.
The unknown coefficient $A(s)$ is determined from the boundary condition (\ref{sph}b):
\begin{align}
\frac{  \kappa_0}{D} A(s)F(\eta R) &=A (s) \eta F'(\eta R) 
   +\partial_rG(R,s|r_0),
\label{CB3}
\end{align}
with
\begin{eqnarray}
\label{boll}
\partial_rG(R,s|r_0)= \frac{1}{D\Omega_dR^{d-1}} \frac{F (\eta r_0)}{F (\eta R)}.
\end{eqnarray}
Rearranging (\ref{CB3}) shows that 
\begin{eqnarray}
\label{AA}
A(s)=\frac{\partial_rG(R,s|r_0) }{\kappa_0F (\eta R)/D-\eta F'(\eta R)} .
\end{eqnarray}

\begin{figure}[t!]
  \centering
  \includegraphics[width=6cm]{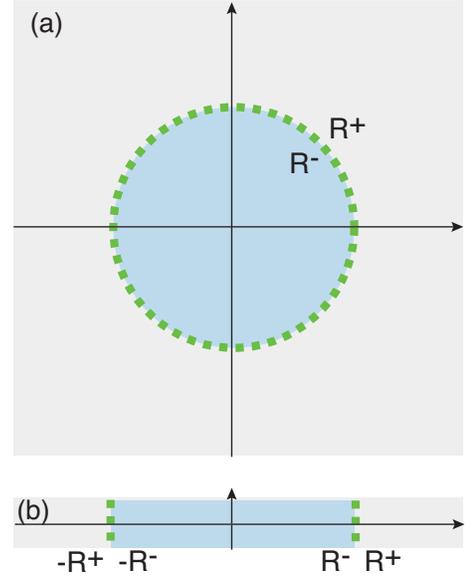}
  \caption{Diffusion through a spherically symmetric semipermeable interface in $\R^d$. (a) For $d=2$ the interface is a circle of radius $R$ (b) For $d=1$ there exist two semipermeable barriers at $r=\pm R$ and a totally reflecting barrier at $r=0^{\pm}$. The solution is reflection symmetric about $r=0$.}
  \label{fig6}
\end{figure}

A similar analysis can be carried out for $\q(r,s|r_0)$ and we find that
 \begin{eqnarray}
    \q(r, s|r_0) = \overline{A}(s)\overline{F}(\eta r) + \overline{G}(r, s|r_0), \,r\leq R,\nonumber \\
    \label{qqir}
\end{eqnarray}
with $\overline{F}(x)=x^{\nu}(c_1I_{\nu}(x)+c_2K_{\nu}(x))$,
\begin{align}
\label{GGbar}
 &  \overline{G}(r, s| r_0) = -\frac{ (rr_0)^\nu }{D\Omega_d}\\
   &\qquad \times \frac{[I_{\nu}(\eta r_>)K_{\nu}(\eta R)-I_{\nu}(\eta R)K_{\nu}(\eta r_>)]I_{\nu}(\eta r_<)}{I_{\nu}(\eta R)},\nonumber
\end{align}
\begin{eqnarray}
\label{bollbar}
-\left . D\frac{d}{dr}\overline{G}(r,s|r_0) \right |_{r=R}= \frac{1}{\Omega_dR^{d-1}} \frac{\overline{F}(\eta r_0)}{\overline{F} (\eta R)},
\end{eqnarray}
and
\begin{eqnarray}
\label{AAbar}
\overline{A}(s)=-\frac{\partial_r\overline{G}(R,s|r_0) }{\kappa_0\overline{F} (\eta R)/D+\eta \overline{F} '(\eta R)}.
\end{eqnarray}
The coefficients $c_1,c_2$ in the definition of $\overline{F}(x)$ depend on the value of $\nu=1-d/2$, and are determined by requiring that the solution remain bounded as $x\rightarrow 0$ and by any symmetries. In particular, for $d=1,2,3$, we have $\nu=1/2,0,-1/2$, respectively, and
\begin{equation} 
\label{Fx}
\overline{F}(x)=\left \{ \begin{array}{cc}\sinh( x) & d=1,\\ \\ I_0(x) & d=2,  \\ \\ \frac{\displaystyle \sinh(  x)}{\displaystyle x} & d=3. \end{array} \right .
\end{equation}
The cases $d=1,3$ follow from the identities
\begin{subequations}\begin{align}
 I_{-1/2}(z) &= \sqrt{\frac{2}{\pi z}}\cosh{(z)},\\
    K_{1/2}(z) &= K_{-1/2}(z) = \sqrt{\frac{\pi}{2 z}}e^{-z},\\
    I_{1/2}(z) &= \sqrt{\frac{2}{\pi z}}\sinh{(z)}.
\end{align}
\end{subequations}
In Fig. \ref{fig6} we show how the one-dimensional case is equivalent to the problem considered in Sect. III, see Fig. \ref{fig2}, with $L'=R$ and $L\rightarrow \infty$. More precisely, spherical symmetry implies that the one-dimensional system is reflection symmetric about $r=0$, which means that there is no flux through the origin.
In other words, we can treat $r=0$ as a totally reflecting barrier. Hence, we can treat diffusion to the right and left of this barrier as independent BMs involving a semipermeable barrier at a distance $R$ from $r=0$.

Finally, substituting Eqs. (\ref{qir}) and (\ref{qqir}) back into Eqs. (\ref{sprenewal}) and noting that $G$ and $\overline{G}$ vanish on the boundary, we have
\begin{subequations}
  \label{spphol}
  \begin{align}
 \widetilde{\rho}(r,s)&= \p(r,s)+\frac{\kappa(R)}{2}A(s)F (\eta r)\Sigma_{\rho}(R,s), \\
 \widetilde{\rho}(r,s)&= \q(r,s)-\frac{\kappa(R)}{2}\overline{A}(s)\overline{F}(\eta r)\Sigma_{\rho}(R,s)
  \end{align}
  \end{subequations} 
for $r\geq R^+$ and $r<R^-$, respectively, and with
\begin{align}
  \label{spam}
\Sigma_{\rho}(R,s)= \frac{ \p(R^+,s)+\q(R^-,s)}{1- \frac{\kappa(R)}{2}[A(s)F (\eta R)+\overline{A}(s)\overline{F} (\eta R))]} . 
 \end{align}

\setcounter{equation}{0}

\section{Encounter-based model of snapping out BM}

As we have already highlighted, one of the advantages of the renewal approach is that it provides a relatively simple framework for developing more general probabilistic models of diffusion through semi-permeable membranes. In our previous paper \cite{Bressloff22p}, we illustrated this in the case of one-dimensional diffusion by considering the effects of (i) stochastic resetting and (ii) modifying the rule for killing each round of partially reflected BM. Here we show how to incorporate the latter into the example of the spherically symmetric interface analyzed in Sect. IVB. The basic idea is to combine
 snapping out BM with the encounter-based model of diffusion-mediated surface absorption \cite{Grebenkov20,Grebenkov22,Bressloff22,Bressloff22a}. This means that each round of partially reflected BM is killed when its local time $\ell_t^{\pm}$ on $\partial \calM^{\pm}$ is greater than an independent random variable $\widehat{\ell}$ with a nonexponential distribution $\Psi(\ell)=\P[\widehat{\ell} > \ell]$. Following Ref. \cite{Bressloff22p} we construct a first rather than a last renewal equation. We add a subscript $\Psi$ to all probability densities in order to indicate the fact we are considering a general distribution threshold distribution $\Psi(\ell)$. 
 
 \subsection{First renewal equation}
 
 The spherically symmetric first renewal equation takes the form
 \begin{widetext}
 \begin{subequations}
 \label{Frenewal}
  \begin{align}
 \rho_{\Psi}(r,t)&=p_{\Psi}(r,t)+\frac{1}{2}\int_0^t [\rho_{\Psi}(r,t-\tau |R^+) +\rho_{\Psi}(r,t-\tau|R^- )]f_{\Psi}(\tau) d\tau ,\, R^+\leq r <\infty,\\
  \rho_{\Psi}(r,t)&=q_{\Psi}(r,t)+\frac{1}{2}\int_0^t [\rho_{\Psi}(r,t-\tau |R^+) +\rho_{\Psi}(r,t-\tau|R^- )]f_{\Psi}(\tau) d\tau , \ 0\leq r \leq R^- . \end{align}
  \end{subequations}
   \end{widetext}
 The first terms on the right-hand sides of Eqs. (\ref{Frenewal}a,b) represent all sample trajectories that have never been absorbed by the barrier at $r=R^{\pm}$ up to time $t$. The corresponding integrals sum over all trajectories that were first absorbed (stopped) at time $\tau$ somewhere on the boundary $ \partial \calM^{\pm}=\{\x,|\x|=R^{\pm}\}$ and then with probability $1/2$ entered the domain $\overline{\calM}$ or its complement, depending on the value of $r$, after which an arbitrary number of switches can occur before reaching $r$ at time $t$. The probability that the first stopping event occurred in the time interval $(\tau,\tau+d\tau)$ is $f_{\Psi}(\tau) d\tau$, 
 where $f_{\Psi}(\tau)$ is the FPT for absorption. Introducing the survival probability
 \begin{equation}
 Q_{\Psi}(t)= \int_{\calM^c} p_{\Psi}(\x,t)d\x+\int_{\calM}q_{\Psi}(\x,t)d\x,
 \end{equation}
 we have 
 \begin{align}
 f_{\Psi}(t)&=-dQ_{\Psi}(t)/dt\\
 &=-D\int_{\calM^c} \frac{\partial p_{\Psi}(\x,t)}{\partial t}d\x-D\int_{\calM}\frac{\partial q_{\Psi}(\x,t)}{\partial t}d\x\nonumber \\
 &=-D\int_{\calM^c} \nabla^2 p_{\Psi}(\x,t)d\x-D\int_{\calM}\nabla^2 q_{\Psi}(\x,t)d\x\nonumber \\
 &=D\int_{\partial \calM^+} \nabla p_{\Psi}(\z,t)\cdot \n d\z\nonumber \\
 &\qquad -D\int_{\partial \calM^-}\nabla q_{\Psi}(\z,t)\cdot \n d\z\nonumber \\
 &=D\Omega_d R^{d-1}[\partial_r p_{\Psi}(R,t)-\partial_r q_{\Psi}(R,t)].
 \end{align}
 The last line follows from spherical symmetry.
 
 Laplace transforming the renewal Eq. (\ref{Frenewal}) with respect to time $t$ gives
\begin{subequations}
 \label{Frenewal2}
  \begin{align}
  &\widetilde{\rho}_{\Psi}(r,s)=\p_{\Psi}(r,s)\\
 &\qquad +\frac{1}{2}  [\widetilde{\rho}_{\Psi}(r,s |R^+) +\widetilde{\rho}_{\Psi}(r,s|R^- )]\widetilde{f}_{\Psi}(s) , \, r\geq R^+, \nonumber \\
  &\widetilde{\rho}_{\Psi}(r,s)=\q_{\Psi}(r,s)\\
 &\qquad +\frac{1}{2}  [\widetilde{\rho}_{\Psi}(r,s |R^+) +\widetilde{\rho}_{\Psi}(r,s|R^- )]\widetilde{f}_{\Psi}(s) , \, r\leq R^-, \nonumber \end{align}
 \end{subequations}
 with 
 \begin{align}
 \widetilde{f}_{\Psi}(s)&=1-s\widetilde{Q}_{\Psi}(s)\nonumber \\
 &=D\Omega_dR^{d-1}[\partial_r\p_{\Psi}(R,s)-\partial_r\q_{\Psi}(R,s)].
 \label{fpsi}
 \end{align}
 In order to determine the factor $\widetilde{\rho}_{\Psi}(r,s |R^+)+\widetilde{\rho}_{\Psi}(r,s |R^-)$ we set $g(r_0)=[\delta(r_0-R^+) +\delta(r_0-R^-)]/2$ in Eq. (\ref{Frenewal2}). This yields
   \begin{align}
&\widetilde{\rho}_{\Psi}(r,s |R^+)+\widetilde{\rho}_{\Psi}(r,s |R^-)=\p_{\Psi}(r,s|R)+\q_{\Psi}(r,s|R) \nonumber \\
&\quad +[\widetilde{\rho}_{\Psi}(r,s |R^+)+\widetilde{\rho}_{\Psi}(r,s |R^-) ]\widetilde{f}_{\Psi}(R,s),
\label{rrr}
 \end{align}
 where $ \widetilde{f}_{\Psi}(R,s)=1-s\widetilde{Q}_{\Psi}(R,s)$ with
 \begin{align}
  \label{Qpsi}
&\widetilde{Q}_{\Psi}(R,s)\\
&=\frac{D\Omega_d R^{d-1}}{2}\lim_{\epsilon\rightarrow 0^+}  [\partial_r p_{\Psi}(R,t|R+\epsilon)-\partial_r q_{\Psi}(R,t|R-\epsilon)] .\nonumber 
\end{align} 
 Note that $\p_{\Psi}(r,s|R)=0$ for $r<R$ and $\q_{\Psi}(r,s|R)=0$ for $r>R$. Rearranging Eq. (\ref{rrr}) leads to the result
    \begin{align*}
\widetilde{\rho}_{\Psi}(r,s |R^+)+\widetilde{\rho}_{\Psi}(r,s |R^-)= \frac{\p_{\Psi}(r,s|R)+\q_{\Psi}(r,s|R)}{s\widetilde{Q}_{\Psi}(R,s)}.
 \end{align*}
   Substituting back into Eq. (\ref{Frenewal2}) yields the explicit solution
  \begin{subequations}
    \label{Frenewal3}
 \begin{align}
& \widetilde{\rho}_{\Psi}(r,s)  = \p_{\Psi}(r,s)+ \frac{1-s\Q_{\Psi}(s)}{2s\widetilde{Q}_{\Psi}(R,s)}\p_{\Psi}(r,s|R),\ r >R\\
& \widetilde{\rho}_{\Psi}(r,s)  = \q_{\Psi}(r,s)+ \frac{1-s\Q_{\Psi}(s)}{2s\widetilde{Q}_{\Psi}(R,s)} \q_{\Psi}(r,s|R), r <R.
 \end{align}
 \end{subequations}
 
 \subsection{Boundary conditions at the interface}

We would like to determine the boundary conditions for $\widetilde{\rho}_{\Psi}(r,s)$ at the interface. In order to proceed further, we use the following general results from studies of encounter-based models \cite{Grebenkov20,Grebenkov22,Bressloff22,Bressloff22a}. Consider partially reflected BM in a bounded domain $\calM$. Let $p(\x,z,t|\x_0)$ be the solution of the corresponding FP equation for constant absorption rate $\kappa_0=zD$ on $\partial \calM$. Then
\begin{subequations}
\label{ppsi0}
\begin{align}
p_{\Psi}(\x,t|\x_0)&=\int_0^{\infty} \Psi(\ell){\mathcal L}_{\ell}^{-1}p(\x,z,t|\x_0)d\ell,\\
-\nabla p_{\Psi}(\y,t|\x_0)\cdot \n &= \int_0^{\infty} \psi(\ell){\mathcal L}_{\ell}^{-1}p(\y,z,t|\x_0)d\ell
\end{align}
\end{subequations}
for $\x,\x_0\in \calM$ and $\y\in \partial \calM$,
with $z$ treated as the Laplace variable conjugate to $\ell$, and $\psi(\ell)=-\Psi'(\ell)$.
In the specific case of a spherically symmetric interface, these results imply that 
\begin{subequations}
\label{ppsi}
\begin{align}
p_{\Psi}(r,t|r_0)&=\int_0^{\infty} \Psi(\ell){\mathcal L}_{\ell}^{-1}p(r,z,t|r_0)d\ell,\\
\partial_r p_{\Psi}(R,t|r_0)  &= \int_0^{\infty} \psi(\ell){\mathcal L}_{\ell}^{-1}p(R,z,t|r_0)d\ell
\end{align}
\end{subequations}
for $r,r_0> R$
and
\begin{subequations}
\label{qpsi}
\begin{align}
q_{\Psi}(r,t|r_0)&=\int_0^{\infty} \Psi(\ell){\mathcal L}_{\ell}^{-1}q(r,z,t|r_0)d\ell,\\
-\partial_r q_{\Psi}(R,t|r_0)  &= \int_0^{\infty} \psi(\ell){\mathcal L}_{\ell}^{-1}q(R,z,t|r_0)d\ell
\end{align}
\end{subequations}
for $r,r_0<R$.

We now calculate the terms $ \p_{\Psi}(r,s|R)$, $ \q_{\Psi}(r,s|R)$ and $s\widetilde{Q}_{\Psi}(R,s)$ appearing in Eqs. (\ref{Frenewal3}).
First, setting $r_0=R$ in Eq. (\ref{qir}) gives
\begin{eqnarray}
    \p(r, z,s|R) = \frac{1}{D\Omega_dR^{d-1}}\frac{F(\eta r) }{zF(\eta R)-\eta F'(\eta R)}   .
    \label{PR}
\end{eqnarray}
Substituting into Eq. (\ref{ppsi}a), we find that
\begin{subequations}
   \label{qirz}
\begin{eqnarray}
    \p_{\Psi}(r,s|R) &=& \frac{\widetilde{\Psi}(\calF(s))}{D\Omega_dR^{d-1}} \frac{F(\eta r)}{F(\eta R)},  \\
      \calF(s)&=&-\frac{\eta F'(\eta R)}{F(\eta R)}.
\end{eqnarray}
\end{subequations}
Similarly, setting $r_0=R$ in Eq. (\ref{qqir}), we have
\begin{eqnarray}
    \q(r, z,s|R) = \frac{1}{D\Omega_dR^{d-1}} \frac{\overline{F}(\eta r) }{z\overline{F}(\eta R)+\eta \overline{F}'(\eta R)}  ,
    \label{QR}
\end{eqnarray}
so that from Eq, (\ref{qpsi}a)
\begin{subequations}
   \label{qqirz}
\begin{eqnarray}
    \q_{\Psi}(r,s|R) &=& \frac{\widetilde{\Psi}(\overline{\calF}(s))}{D\Omega_dR^{d-1}} \frac{\overline{F}(\eta r)}{\overline{F}(\eta R)},\\
    \overline{\calF}(s)&=&\frac{\eta \overline{F}' (\eta R)}{\overline{F}(\eta R)}.
  \end{eqnarray}
\end{subequations}
Finally, we determine $s\Q_{\Psi}(R,s)$ by combining Eqs. (\ref{Qpsi}), (\ref{ppsi}b), (\ref{PR}) and (\ref{QR}):
\begin{widetext}
\begin{align}
&s\Q_{\Psi}(R,s)=1-\frac{D\Omega_d R^{d-1}}{2}\int_0^{\infty} \psi(\ell){\mathcal L}^{-1}[\p(R,s,z|R)+\q(R,s,z|R)]d\ell=1-\frac{1}{2}\left [ \widetilde{\psi}(\calF(s))+\widetilde{\psi}(\overline{\calF}(s))\right ].
\end{align}

Determining the boundary conditions at the interface requires differentiating both sides of Eqs. (\ref{Frenewal3}) with respect to $r$ and setting $r=R^{\pm}$. This yields terms of the form $\partial_r\p_{\Psi}(R,s |R) $ and $\partial_r\q_{\Psi}(R,s |R) $. Since the initial state is on the boundary we cannot simply set $r_0=R$ in Eqs. (\ref{ppsi}b) and (\ref{qpsi}b). Instead, we differentiate Eqs. (\ref{qirz}) and (\ref{qqirz}) directly:
\begin{subequations}
\begin{align}
&\partial_r\p_{\Psi}(R,s |R) =\frac{1}{D\Omega_dR^{d-1}}\sqrt{\frac{s}{D}}  \widetilde{\Psi}(\calF(s))\frac{F'(\sqrt{s/D}R)}{F(\sqrt{s/D}R)}=-\frac{1}{D\Omega_dR^{d-1}} \widetilde{\Psi}(\calF(s))\calF(s),\\
& \partial_r\q_{\Psi}(R,s |R)=\frac{1}{D\Omega_dR^{d-1}}\sqrt{\frac{s}{D}}\widetilde{\Psi}(\overline{\calF}(s))\frac{\overline{F}'(\sqrt{s/D}R)}{\overline{F}(\sqrt{s/D}R)}=\frac{1}{D\Omega_dR^{d-1}}\widetilde{\Psi}(\overline{\calF}(s))\overline{\calF}(s).
\end{align}
\end{subequations}
Differentiating Eqs. (\ref{Frenewal3}) with respect to $r$, setting $r=R^{\pm}$ and subtracting the results
 \begin{align}
 \label{enc1}
& \partial_r\widetilde{\rho}_{\Psi}(R^+,s) - \partial_x\widetilde{\rho}_{\Psi}(R^-,s) = \partial_r\p_{\Psi}(R,s)- \partial_r\q_{\Psi}(R,s)- \frac{1}{D\Omega_dR^{d-1}}\frac{\widetilde{\Psi}(\calF(s))\calF(s)+\widetilde{\Psi}(\overline{\calF}(s))\overline{\calF}(s)}{2-\widetilde{\psi}(\calF(s))-\widetilde{\psi}(\overline{\calF}(s))} \widetilde{f}_{\Psi}(s). \end{align}
Since $\widetilde{\psi}(\calF)\calF=1-\widetilde{\psi}(\calF)$ etc., we deduce from Eq. (\ref{fpsi}) that
\begin{align}
  \partial_r\widetilde{\rho}_{\Psi}(R^+,s) - \partial_x\widetilde{\rho}_{\Psi}(R^-,s) = \partial_r\p_{\Psi}(R,s)- \partial_r\q_{\Psi}(R,s)-\frac{\widetilde{f}_{\Psi}(s)}{D\Omega_dR^{d-1}}=0.
  \end{align}
  Hence, the probability flux is continuous across the interface $\partial \calM$. Next, differentiating Eqs. (\ref{Frenewal3}) with respect to $r$, setting $r=R^{\pm}$ and adding the results gives
 \begin{align}
& 2\partial_r\widetilde{\rho}_{\Psi}(R^{\pm},s) = \partial_r\p_{\Psi}(R,s)+ \partial_r\q_{\Psi}(R,s)-  \frac{\widetilde{\Psi}(\calF(s))\calF(s)-\widetilde{\Psi}(\overline{\calF}(s))\overline{\calF}(s)}{D\Omega_dR^{d-1}} \frac{\widetilde{f}_{\Psi}(s)}{2s\widetilde{Q}(R,s)}\nonumber\\
&=\partial_r\p_{\Psi}(R,s)+ \partial_r\q_{\Psi}(R,s)+ \frac{\widetilde{\psi}(\calF(s))-\widetilde{\psi}(\overline{\calF}(s))}{D\Omega_dR^{d-1}}  \frac{\widetilde{f}_{\Psi}(s)}{2s\widetilde{Q}(R,s)}\nonumber \\
&=\partial_r\p_{\Psi}(R,s)+ \partial_r\q_{\Psi}(R,s)+\left [ \frac{\widetilde{\psi}(\calF(s))}{\widetilde{\Psi}(\calF(s))}\p_{\Psi}(R,s|R)-\frac{\widetilde{\psi}(\overline{\calF}(s))}{ \widetilde{\Psi}(\overline{\calF}(s))} \q_{\Psi}(R,s|R) \right ] \frac{\widetilde{f}_{\Psi}(s)}{2s\widetilde{Q}(R,s)}.
 \end{align}
Finally,
\begin{align}
 \partial_r\p_{\Psi}(R,s)+ \partial_r\q_{\Psi}(R,s)&= \int_0^{\infty} \psi(\ell)\bigg \{ \int_R^{\infty}{\mathcal L}_{\ell}^{-1}\p(R,z,s|r_0)g(r_0)dr_0 -  \int_0^{R}{\mathcal L}_{\ell}^{-1}\q (R,z,s|r_0)g(r_0)]dr_0\bigg \}d\ell\nonumber\\
&=\frac{1}{D\Omega_dR^{d-1}} \int_R^{\infty}\widetilde{\psi}(\calF(s))  \frac{F(\sqrt{s/D}R)}{F(\sqrt{s/D}r_0)}g(r_0)dr_0 -  \int_0^{R}\widetilde{\psi}(\overline{\calF}(s))\frac{\overline{F}(\sqrt{s/D}R)}{\overline{F}(\sqrt{s/D}r_0)}g(r_0)dr_0\nonumber \\
&=\frac{\widetilde{\psi}(\calF(s)) }{\widetilde{\Psi}(\calF(s)) }\p_{\Psi}(R,s)-\frac{\widetilde{\psi}(\overline{\calF}(s))}{\widetilde{\Psi}(\overline{\calF}(s))}\q_{\Psi}(R,s).
\end{align}
Hence
\begin{align}
2\partial_r\widetilde{\rho}_{\Psi}(R^{\pm},s) &= \frac{\widetilde{\psi}(\calF(s)) }{\widetilde{\Psi}(\calF(s)) } \widetilde{\rho}_{\Psi}(R^+,s)-
\frac{\widetilde{\psi}(\overline{\calF}(s))}{\widetilde{\Psi}(\overline{\calF}(s))}\widetilde{\rho}_{\Psi}(R^-,s).
\label{fin}
\end{align}

\end{widetext}

Introducing the Laplace transforms
\begin{equation}
\kappa_+(s)= D\frac{\widetilde{\psi}(\calF(s)) }{\widetilde{\Psi}(\calF(s)) },\quad \kappa_-(s)=D\frac{\widetilde{\psi}(\overline{\calF}(s))}{\widetilde{\Psi}(\overline{\calF}(s))},
\label{skap}
\end{equation}
we can rewrite the boundary condition (\ref{fin}) in the more suggestive form
\begin{align}
2D\partial_r\widetilde{\rho}_{\Psi}(R^{\pm},s) &=\kappa_+(s)\widetilde{\rho}_{\Psi}(R^+,s)-
\kappa_-(s)\widetilde{\rho}_{\Psi}(R^-,s).
\label{fin2}
\end{align}
This can be inverted using the convolution theorem for Laplace transforms to yield the following result:
\begin{align}
\label{final}
2D\partial_r\rho_{\Psi}(R^{\pm},t)&=\int_0^{\infty} [\kappa_+(t-\tau)\rho(R^+,\tau)\nonumber \\
&\qquad -\kappa_-(t-\tau)\rho(R^-,\tau)]d\tau .
\end{align}
  That is, the inward flux into the sphere is determined by an asymmetric pair of time-dependent permeabilities with memory. This asymmetry occurs even though the non-Markovian absorption process on either side of the interface is the same. In the special case of an exponential distribution, we have $\widetilde{\psi}(s)=\kappa_0\widetilde{\Psi}(s)/D$ for all $s$ so that $\kappa_{\pm}(s)=\kappa_0/D$ and $\kappa(t-\tau)=(\kappa_0/D) \delta(t-\tau)$. We thus recover the classical permeable boundary condition. Note that a boundary condition of the form (\ref{final}) has recently been considered within the context of a subdiffusion model, in which anomalous behavior is generated by a thin membrane with a non-exponential waiting time density for the particle sojourn time within the membrane \cite{Kos21}.

\subsection{Analysis of permeability functions}

For the sake of illustration, suppose that $\psi(\ell) $ is given by the gamma distribution:
\begin{equation}
\label{psigam}
\psi(\ell)=\frac{\gamma(\gamma \ell)^{\mu-1}\e^{-\gamma \ell}}{\Gamma(\mu)}, \mu >0,
\end{equation}
where $\Gamma(\mu)$ is the gamma function. The corresponding Laplace transforms are
\begin{equation}
\widetilde{\psi} (z)=\left (\frac{\gamma}{\gamma+z}\right )^{\mu},\quad \widetilde{\Psi}(z)=\frac{1-\widetilde{\psi}(z)}{z}.
\end{equation}
Here $\gamma$ determines the effective absorption rate. If $\mu=1$ then $\psi$ reduces to the exponential distribution with constant reactivity $\kappa_0 = D\gamma$. The parameter $\mu$ thus characterizes the deviation of $\psi(\ell)$ from the exponential case. If $\mu <1$ ($\mu>1$) then $\psi(\ell)$ decreases more rapidly (slowly) as a function of the local time $\ell$. Substituting the gamma distribution into Eqs. (\ref{skap}) yields
\begin{subequations}
 \begin{align}
 \widetilde{\kappa}_+(s)&=D\frac{\calF(s)\gamma^{\mu}}{(\gamma+\calF(s))^{\mu}-\gamma^{\mu}},\\ \widetilde{\kappa}_-(s)&=D\frac{\overline{\calF}(s)\gamma^{\mu}}{(\gamma+\overline{\calF}(s))^{\mu}-\gamma^{\mu}}.
 \end{align}
 \end{subequations}
 If $\mu=1$ then $\widetilde{\kappa}(s)=\gamma D=\kappa_0$ and $\kappa(\tau)=\kappa_0\delta(\tau)$ as expected. In order to explore an example of a nonexponetial distribution we take $\mu=2$ such that
  \begin{equation}
 \widetilde{\kappa}_+(s)=\frac{\kappa_0}{2}\frac{1}{1+\frac{\dis D}{\dis 2\kappa_0}\calF(s)},\  \widetilde{\kappa}_-(s)=\frac{\kappa_0}{2}\frac{1}{1+\frac{\dis D}{\dis 2\kappa_0}\overline{\calF}(s)}.
 \end{equation}
 
 In the one-dimensional case ($d=1)$ we have, see Eq. (\ref{Fx}),
\begin{equation}
  \calF(s)=\sqrt{\frac{s}{D}}   ,\ \overline{\calF}(s)=\sqrt{\frac{s}{D}}\mbox{tanh}(\sqrt{s/D}R),
  \end{equation}
  and
  \begin{subequations}
  \label{kappm}
  \begin{align}
   \widetilde{\kappa}_+(s)&=\frac{\kappa_0}{2}\frac{1}{1+\frac{\dis \sqrt{sD}}{\dis 2\kappa_0}},\\  \widetilde{\kappa}_-(s)&=\frac{\kappa_0}{2} \frac{1}{1+\frac{\dis \sqrt{sD}}{\dis 2\kappa_0} \mbox{tanh}(\sqrt{s/D}R)},
   \end{align}
   \end{subequations}
   with $\kappa_0=\gamma D$. We first consider the permeability on the right-hand side of the barrier at $x=R$, see Fig. \ref{fig6}(b).
 The function $\widetilde{\kappa}_+(s)$ is identical to the permeability function on either side of a semipermeable barrier in $\R$ \cite{Bressloff22p}, and has the explicit inverse
  \begin{align}
  \label{kaptp}
 \kappa_+(\tau)=\frac{\kappa_0^2}{\sqrt{D}}\left [\frac{1}{\sqrt{\pi \tau}}-\frac{2\kappa_0}{\sqrt{D}}\e^{4\kappa_0^2\tau/D} \mbox{erfc}(2\kappa_0\sqrt{\tau/D})\right ],
 \end{align}
 where $\mbox{erfc}(x)=(2/\sqrt{\pi})\int_x^{\infty} \e^{-y^2}dy$ is the complementary error function. The permeability $\kappa_+(\tau)$ is a monotonically decreasing function of time with $\kappa_+(t)\rightarrow 0$ as $t\rightarrow \infty$. In addition, it is a heavy-tailed distribution with infinite moments. The latter follows from the large-$t$ behavior of $\kappa_+(\tau)$, which can be determined by performing a small-$s$ expansion and using \cite{Redner01}
\begin{equation}
\int_0^{\infty} \e^{-st}\frac{t^{\alpha-1}}{\Gamma(\alpha)}dt =s^{-\alpha}.
\end{equation}
Although this formula only holds for $\mbox{Re}(\alpha)>0$, it can be extended in the complex $\alpha$-plane (excluding $\alpha =0,-1,-2,\ldots$) using the theory of distributions; the resulting singular terms can then be ignored when considering the large-$t$ behavior. Taylor expanding $\widetilde{\kappa}_+(s)$ as a function of $s$, we find that
\begin{align}
 \widetilde{\kappa}_+(s)&= \frac{\kappa_0}{2}\frac{1-\sqrt{sD}/2\kappa_0}{1-sD/4\kappa_0^2} \\
 &\sim \frac{\kappa_0}{2[1-sD/4\kappa_0^2]}+\frac{\kappa_0}{2}\left [-\frac{\sqrt{sD}}{2\kappa_0}-\frac{(sD)^{3/2}}{8\kappa_0^3}\right ]\nonumber
 \end{align}
 as $s\rightarrow 0$.
 Hence,
 \begin{equation}
 \kappa(t)\sim \sqrt{\frac{D}{\pi}} \frac{1}{8t^{3/2}} ,\ t \rightarrow \infty.
 \end{equation}
This is consistent with asymptotically expanding $\mbox{erfc}(x)$ in Eq. (\ref{kaptp}) using the formula
 \begin{equation}
 \mbox{erfc}(x)\sim \frac{1}{\sqrt{\pi}}\e^{-x^2} \sum_{k=0}^{\infty} (-1)^k\frac{(2k)!}{2^{2k}k!}\frac{1}{x^{2k+1}},
 \end{equation}
 whose rate of decay depends on $\kappa_0$ and $D$. 
 
 Turning to the permeability on the left-hand side of the barrier, $x=R^-$, we note from Eq. (\ref{kappm}b) that $\widetilde{\kappa}_-(s)$ has an infinite set of poles in the negative half of the complex-$s$ plane. These are determined from the zeros of the function $f(x)=1+(D/2\kappa_0)x\tanh(x)$ with $x=\sqrt{s/D}R$. The zeros also correspond to the discrete spectrum of the diffusion operator in the bounded interval $[0,R]$. The smallest  eigenvalue is real and determines the exponential rate of decay in the large-$t$ limit.
   
 In the three-dimensional case ($d=3$)
 \begin{equation}
  \calF(s)=\sqrt{\frac{s}{D}}   ,\ \overline{\calF}(s)=\sqrt{\frac{s}{D}}\left [\mbox{coth}(\sqrt{s/D}R)-\frac{1}{\sqrt{s/D}R}\right ],
  \end{equation}
so that $\widetilde{\kappa}_+(s)$ is the same as for $d=1$, whereas
  \begin{align}
   \widetilde{\kappa}_-(s)&=\frac{\kappa_0}{2}\frac{1 }{1+\frac{\dis \sqrt{sD}}{\dis 2\kappa_0} \left [\mbox{coth}(\sqrt{s/D}R) -\frac{\dis 1}{\dis \sqrt{s/D}R}\right ]}.
   \end{align}
Again the permeability $\kappa_-(t)$ decays exponentially for large $t$, except now the decay rate is determined by smallest negative zero of the function $f(x)=1+(D/2\kappa_0)(\mbox{coth}(x)-x^{-1})$. Finally, when $d=2$ we have
 \begin{subequations}
 \begin{align}
  \calF(s)&= -\sqrt{\frac{s}{D}}\frac{  K'_{0}(\sqrt{s/D} R)}{K_{0}(\sqrt{s/D} R)},\\
    \overline{\calF}(s)&=\sqrt{\frac{s}{D}}\frac{  I'_{0}(\sqrt{{s}/{D}} R)}{I_{0}(\sqrt{{s}/{D}} R)}.    \end{align}
  \end{subequations} 
  The zero-order modified Bessel functions have the following small-$s$ expansions:
  \begin{align}
 I_0(x)&=1+\frac{x^2}{4}+O(x^4),\\
  K_0(x)&=-[\log(x/2)+\gamma_e]I_0(x)+\frac{x^2}{4}+O(x^4),
 \end{align}
 where $\gamma_e$ is Euler's constant, that is, $\gamma_e\approx 0.5772$. Hence,
 \begin{align}
 \calF(s)\sim -\frac{1}{R\log(\sqrt{s/D}R/2)}
  \end{align}
  for $s\rightarrow 0$
It follows that
\begin{align}
\widetilde{\kappa}_+(s)\sim  \frac{\kappa_0}{2}\left [1+\frac{D}{\kappa_0 R \log s}\right ],
\end{align}
and thus
\begin{equation}
\kappa_+(t)\sim \frac{D}{2R}\frac{1}{t(\log t)^2}.
\end{equation}
On the other hand, $\kappa_-(t)$ decays exponentially at a rate determined by the leading order zero of $I_0(x)$. Finally, note that another mechanism for generating power law behavior would be to consider a heavy-tailed distribution $\psi(\ell)$ such as a Pareto-II (Lomax) distribution \cite{Grebenkov20}.
 
  \bigskip

  \section{Conclusion} 
  
In this paper we established the equivalence between snapping out BM and single particle diffusion through a semipermeable interface for several simple geometries. Examples included an asymmetric barrier in a one-dimensional bounded domain, and a higher-dimensional closed membrane in $\R^d$. In each case we derived a renewal equation relating the full probability density to the probability densities of partially reflected BM on either side of the interface. The renewal equations were solved using a combination of Laplace transforms and Green's function methods. One of the potential advantages of the renewal approach is that it provides a  probabilistic framework for developing more general models of semipermeable membranes. We illustrated this by considering an encounter-based model of absorption on either side of a spherically symmetric interface. (Absorption is the mechanism that kills each round of partially reflected BM.)  In particular, we showed that non-Markovian models of absorption generate an asymmetric time-dependent permeability that tends to be heavy-tailed. 

Our formulation in terms of renewal equations also provides an alternative method for solving classical boundary value problems in the presence of a semipermeable interface, at least in the Laplace domain. We considered the particular example of a one-dimensional first passage time problem, in which a semipermeable barrier was placed between a reflecting boundary and an absorbing boundary. The MFPT was calculated in terms of the Laplace transformed flux through the absorbing boundary. This raises a more general issue, namely, can the renewal approach simplify the analysis of certain boundary value problems in more complex media containing multiple interfaces and heterogeneous diffusivities. This would require developing efficient numerical schemes for solving the renewal equation directly or for implementing snapping out BM. There has been considerable recent interest in finding hybrid analytical/numerical methods for solving the diffusion equation in multilayered media \cite{Grebenkov10,Hahn12,Lejay12,Carr16,Moutal19,Farago20,Alemany22}. 

Finally, another possible application of the renewal approach is to incorporate a stochastic resetting protocol, see the review \cite{Evans20}. In our previous paper, we analyzed stochastic resetting in the case of diffusion through a semipermeable barrier in $\R$ and studied the relaxation to a nonequilibrium stationary state (NESS) in the large time limit \cite{Bressloff22p}. One of the novel features arising from the presence of a semipermeable interface is that it is natural to exclude resetting paths that cross the interface, which can lead to a space-dependent form of resetting.


\vfill


  
\begin{thebibliography}{9}

\bibitem{Kedem58} O. Kedem and A. Katchalsky, Thermodynamic analysis of the permeability of biological membrane to non-electrolytes. {Biochim. Biophys. Acta} {\bf 27} 229-246 (1958).

\bibitem{Kedem62} A. Katchalsky and O. Kedem, Thermodynamics of Flow Processes in Biological Systems. Biophys. J. {\bf 2}  53-78 (1962).

\bibitem{Kargol96} A. Kargol, M. Kargol and S. Przestalski, The Kedem-Katchalsky equations as applied for describing substance transport across biological membranes, Cell. Mol. Biol. Lett. {\bf 2} 117-124 (1996) 

\bibitem{Aho16} V. Aho, K. Mattila, T. K\"{u}hn, P. Kek\"{a}l\"{a}inen, O. Pulkkine, R. B. Minussi,
M. Vihinen-Ranta and J. Timonen, Diffusion through thin membranes: Modeling across scales. Phy. Rev. E {\bf 93} 043309 (2016)


\bibitem{Nik21} V. Nikonenko and
N. Pismenskaya (Eds.). Ion and Molecule Transport in Membrane Systems (special issue). Int.
J. Mol. Sci. {\bf 22} 3556 (2021).

\bibitem{Li10} D. Li and H. Wang, Recent developments in reverse osmosis desalination membranes J. Mater. Chem. {\bf 20} 4551 (2010).

\bibitem{Rubinstein21} I. Rubinstein, A. Schur and B. Zaltzman, Artifact of ``breakthrough'' osmosis: Comment on the
local Spiegler-Kedem-Katchalsky equations with constant coefficients. Sci. Rep. {\bf 11} 5051 (2021).

\bibitem{Powles92} J. G. Powles, M. Mallett, G. Rickayzen and W. Evans, Exact
analytic solutions for diffusion impeded by an infinite array of
partially permeable barriers {Proc. R. Soc. Lond. A} {\bf 436} 391 (1992)


\bibitem{Kenkre08} V. M. Kenkre, L. Giuggiol and Z. Kalay Molecular motion
in cell membranes: analytic study of fence-hindered random
walks { Phys. Rev. E} {\bf 77} 051907 (2008)

\bibitem{Novikov11} D. Novikov, E. Fieremans, J. Jensen and J. A. Helpern, Random walks with barriers. Nat. Phys. {\bf 7} 508-514 (2011)

\bibitem{Kay22} T. Kay and L. Giuggioli, Diffusion through permeable interfaces: Fundamental equations and their application to
first-passage and local time statistics. {Phys. Rev. Res.} {\bf 4} L032039 (2022).



   \bibitem{Ito65} K. Ito and H. P. McKean {\em Diffusion Processes and Their Sample Paths} Springer-Verlag,
Berlin (1965)
  
   \bibitem{Freidlin85} M. Freidlin, {\em Functional Integration and Partial Differential Equations}
Annals of Mathematics Studies, Princeton University Press, Princeton
New Jersey (1985)

\bibitem{Papanicolaou90} V. G. Papanicolaou, {The probabilistic solution of the third boundary
value problem for second order elliptic equations} {Probab. Th. Rel. Fields}
{\bf 87}, 27-77 (1990)

 \bibitem{Milshtein95} G. N. Milshtein, {The solving of boundary value problems by numerical
integration of stochastic equations.} {Math. Comp. Sim.} {\bf 38} 77-85 (1995)


\bibitem{Borodin96} A. N. Borodin and P. Salminen, {\em Handbook of Brownian Motion: Facts and Formulae}
Birkhauser Verlag, Basel-Boston-Berlin (1996).

\bibitem{Grebenkov06} D. S. Grebenkov, Partially Reflected Brownian Motion: A Stochastic
Approach to Transport Phenomena, in ``Focus on Probability Theory'',
Ed. Velle LR pp. 135-169 Hauppauge: Nova Science Publishers (2006)

\bibitem{Lejay16} A. Lejay, The snapping out Brownian motion. {The Annals of Applied Probability}
 {\bf 26} 1727-1742 (2016).
 
 \bibitem{Lejay18} A. Lejay, A Monte Carlo estimation of the mean residence time in cells surrounded by thin layers. Mathematics and Computers in Simulation
{\bf 143} 65-77 (2018)

 \bibitem{Brobowski21} A. Bobrowski, Semigroup-theoretic approach to diffusion in thin layers separated by semi-permeable membranes. J. Evol. Equ. {\bf 21} 1019-1057 (2021). 
 
 \bibitem{Ito63} K. Ito and H. McKean  Brownian motions on a half line {Illinois J.Math.} {\bf 7} 181-231 (1963).
 
 \bibitem{Lejay06} A. Lejay, On the constructions of the skew Brownian motion {Probab. Surv.} {\bf 3} 413-466 (2006).
 

 
\bibitem{Decamps06} M. Decamps, M. Goovaerts and W. Schoutens, Asymmetric skew Bessel processes and their applications to finance, {J. Comput. Appl. Math.} {\bf 186} 130-147 (2006).
 
\bibitem{App11} T. Appuhamillage, V. Bokil, E. Thomann, E. Waymire and B. Wood, Occupation and local times for skew Brownian motion with applications to dispersion across an interface. {Ann. Appl. Probab.} {\bf 21} 183-214 (2011).

\bibitem{Gairat17} A. Gairat and V. Shcherbakov, Density of skew Brownian motion and its functionals with application in finance. {Math. Finance} {\bf 27} 1069-1088 (2017)

\bibitem{Farago18} S. Regev and O. Farago, Application of underdamped Langevin dynamics simulations for the study of diffusion from a drug-eluting stent, Phys. A, Stat. Mech. Appl. {\bf 507} 231-239 (2018).

 \bibitem{Farago20} O. Farago, Algorithms for Brownian dynamics across discontinuities. J. Comput. Phys. {\bf 423} 109802 (2020).


 
 \bibitem{Bressloff22p} P. C. Bressloff, A probabilistic model of diffusion through a semipermeable membrane. arXiv:2209.09176 (2022)
 
 \bibitem{Evans20} M. R. Evans, S. N. Majumdar and G. Schehr, Stochastic resetting and applications. {J. Phys. A: Math. Theor.} {\bf 53} 193001 (2020).
 
 \bibitem{Grebenkov20} D. S. Grebenkov,  {Paradigm shift in diffusion-mediated surface phenomena.} {Phys. Rev. Lett.} {\bf 125}, 078102 (2020)


\bibitem{Grebenkov22} D. S. Grebenkov,  {An encounter-based approach for restricted diffusion with a gradient drift.}  {J. Phys. A.} {\bf 55} 045203 (2022)

\bibitem{Bressloff22} P. C. Bressloff,  Diffusion-mediated absorption by partially reactive targets: Brownian functionals and generalized propagators. {J. Phys. A.} {\bf 55} 205001 (2022)

\bibitem{Bressloff22a} P. C. Bressloff, Spectral theory of diffusion in partially absorbing media. {Proc. R. Soc. A} {\bf 478} 20220319 (2022)


\bibitem{Kos21} T. Kosztolowicz and A. Dutkiewicz, Boundary conditions at a thin membrane for the normal diffusion equation which generate subdiffusion.
Phys. Rev. E {\bf 103} 042131 (2021)



 
 \bibitem{note1}Recall that a continuous stochastic process $\{X_t\, \ t\geq 0\}$ is said to have the Markov property if the conditional probability distribution of future states of the process (conditional on both past and present states) depends only upon the present state, not on the sequence of events that preceded it. That is, for all $t'>t$ we have $\P[X_{t'}\leq x|X_{s},s\leq t]=\P[X_{t'}\leq x|X_{t}]$.
 The {strong Markov property} is similar to the Markov property, except that the ``present'' is defined in terms of a stopping time.



 \bibitem{Redner01} S. Redner. {\em A Guide to First-passage Processes.} Cambridge University Press, Cambridge (2001).

\bibitem{Grebenkov10} D. S. Grebenkov, Pulsed-gradient spin-echo monitoring of restricted diffusion in multilayered structures.
J. Magn. Reson. {\bf 205} 181-195 (2010)


\bibitem{Hahn12} D. W. Hahn and M. N. Ozisik, One-Dimensional Composite Medium, Ch. 10 pp. 393-432. Wiley, Hoboken (2012).

\bibitem{Lejay12} A. Lejay and G. Pichot, Simulating diffusion processes in discontinuous media: a numerical scheme with
constant time steps. J. Comput. Phys. {\bf 231} 7299-7314 (2012)

\bibitem{Carr16} E. Carr and I. Turner, A semi-analytical solution for multilayer diffusion in a composite medium consisting
of a large number of layers. Appl. Math. Model. {\bf 40} 7034-7050 (2016)


\bibitem{Moutal19} N. Moutal and D. S. Grebenkov,Diffusion across semi-permeable
barriers: spectral properties, efficient computation, and applications,
{J. Sci. Comput.} {\bf 81} 1630-1654 (2019).

\bibitem{Alemany22} I. Alemany, J. N. Rose. J Garnier-Brun, A. D. Scott and D. J. Doorly, Random walk diffusion simulations
in semi-permeable layered media with varying diffusivity, Science Reports {\bf 12} 10759 (2022).



\end{thebibliography}
\end{document}